\documentclass[12pt,a4paper]{article}

\usepackage{eurosym}
\usepackage{amsfonts}
\usepackage{amssymb}
\usepackage{amsmath}
\usepackage{amsthm}
\usepackage{amstext}
\usepackage[dvips]{graphicx}
\usepackage{graphics}
\usepackage{lscape}
\usepackage{ulem}
\usepackage[usenames,dvipsnames]{xcolor}
\usepackage[latin1]{inputenc}
\usepackage{epsfig}
\usepackage{soul}
\usepackage{subfig}
\usepackage[]{natbib}
\usepackage[]{enumerate}
\usepackage{booktabs}%

\emergencystretch=\maxdimen
\newtheoremstyle%
 {redthm}%
 {}{}%
 {\color{black}\itshape}
 {}%
 {\color{OrangeRed}\bfseries}%
 {\color{OrangeRed}}%
 { }{}

\newtheoremstyle%
 {orangethm}%
 {}{}%
 {\color{black}\itshape}
 {}%
 {\color{PineGreen}\bfseries}%
 {\color{PineGreen}}%
 { }{}

\theoremstyle{redthm}
\newtheorem{theorem}{Theorem}
\theoremstyle{orangethm}

\theoremstyle{orangethm}

\begin{document}

\title{
The exclusion dilation operator for bilateral claims problems\thanks{I would like to thank Juan D. Moreno-Ternero, Ricardo Mart\'{i}nez and participants at conferences in Sevilla and Vigo for their helpful comments. The usual disclaimer applies. Financial support from grant PID2023-146364NB-I00, funded by MCIU/AEI/10.13039/501100011033 and FSE+.}
}

\author{\textbf{Aitor Calo-Blanco\thanks{Corresponding author at: Departamento de Econom\'{i}a, Universidade da Coru\~{n}a, Campus de Elvi\~{n}a, 15071 A Coru\~{n}a, Spain. E-mail: aitor.calo@udc.es. ORCID ID: 0000-0003-3957-4741.}} \\ Universidade da Coru\~{n}a}

\date{\today}

\maketitle

\begin{abstract}

This paper examines bilateral claims problems with lower and upper exclusion thresholds that determine whether an individual is excluded from initial gains or losses. We introduce the exclusion dilation operator, a method that transforms standard rules into extended rules incorporating exclusion thresholds. The operator first allocates gains and losses with respect to these thresholds and then distributes the remaining resources through a dilation transformation of an underlying standard rule. We axiomatically characterize this operator and examine which standard properties of the theory of fair allocation it preserves. While the operator maintains key properties such as homogeneity and monotonicity, it intentionally violates others, most notably order preservation, to reflect the asymmetries induced by exclusion thresholds. Our approach provides a formal methodology for resource allocation in contexts where symmetry is not appropriate due to legal and policy considerations.

\vspace{0.15cm}


\vspace{0.05cm}


\end{abstract}


\section{Introduction}
\label{sect_intro}

A conflicting claims problem, also known as a rationing problem, involves the allocation of a limited and divisible resource, the endowment, among individuals whose aggregate claims exceed what is available. After the seminal contribution by \cite{Oneill_82_MSS}, a substantial literature has formally characterized allocation rules that satisfy desirable fairness and robustness properties \citep[see][]{Moulin_02_handbook, Thomson_03_MSS, Thomson_15_MSS, Thomson_19_Book}. These models provide solutions to situations involving conflicting interests, including bankruptcy \citep{Aumann_al_85_JET}, taxation \citep[][]{Chambers_al_17_SCWE}, transportation cost division \citep[][]{Estan_al_21_AOR}, sportscast revenue sharing \citep[][]{Bergantinos_al_20_MS, Bergantinos_al_26_SER}, CO$_2$ emissions allocation \citep[][]{Ju_al_21_EE}, and tournament slot assignment \citep[][]{Krumer_al_23_JSE}.

This paper analyzes bilateral claims problems with agents subject to lower and upper exclusion thresholds. These thresholds determine whether an individual is excluded from initial gains or losses, thereby incorporating asymmetries into the allocation process. We introduce a new mechanism -- the \textit{exclusion dilation operator} -- which, once exclusion thresholds are reached, allocates the remaining resources through a dilation transformation of a rule for the domain of problems without exclusions. We show that while this operator preserves fundamental properties (e.g., homogeneity and monotonicity), its inherent structure leads to the violation of others, such as order preservation. Finally, we provide an axiomatic characterization of the operator.

Exclusion thresholds capture meaningful baseline entitlements and priority constraints. A lower exclusion threshold guarantees minimum awards whenever resources permit, while an upper exclusion threshold defers full compensation for the claimant until other agents have reached comparable entitlements. These thresholds are particularly relevant in settings where symmetric treatment is not appropriate due to fairness and policy considerations \citep[e.g.,][]{Moulin_00_Econom, Hokari_al_03_ET, Stovall_14_GEB}. For instance, government claims in bankruptcy proceedings may be treated differently from those of private creditors. Similarly, tax systems may exempt part of the revenue generated from productive activities.

Many conflicting scenarios involve bilateral relationships, such as those between employers and employees, teachers and students, or spouses. In these contexts, decision-makers often establish exclusion thresholds before allocating resources between the parties involved. For example, educational funds may be partitioned to ensure that students can acquire basic course materials before the remaining funds are distributed. In divorce settlements, one spouse may receive specific assets as compensation for career sacrifices made during the marriage. The bilateral problem provides a critical benchmark because different allocation rules may coincide with the same formula in this context. Concede-and-divide, for instance, is the two-claimant version of the Talmud, adjusted proportional, random arrival, minimal overlap, and average-of-awards rules \citep[see][]{Thomson_19_Book}. By focusing on this setting, we can show the precise allocation dynamics resulting from the asymmetries inherent in exclusion thresholds.

Our paper builds on the operator approach for claims problems. \cite{Thomson_al_08_JET} introduced operators as mappings from the space of rules into itself, characterizing the resulting formulas that arise from subjecting a division rule to duality, truncation of excessive claims, and attribution of minimal rights. \cite{Hougaard_al_12_JME} generalized this concept by defining a family of operators involving baseline rationing, composition properties, and lower bounds. We introduce a dilation-based operator, which, rather than truncating claims or imposing fixed baselines, compresses the claims space and rescales the structure of the underlying rule.

The exclusion dilation operator uses a three-stage process to transform standard bilateral allocation rules into extended rules incorporating lower and upper exclusion thresholds. In the first stage, resources are assigned exclusively to the agent not excluded from the initial gains until the lower exclusion threshold is satisfied. In the second stage, once this threshold is met, the remaining endowment is distributed according to a dilated version of a standard rule, where the dilation reflects the contraction of the claims space induced by the two exclusion thresholds. In the final stage, the agent excluded from participating in small windfalls by the upper exclusion threshold is compensated in full, and any residual is allocated to the other agent.

The exclusion dilation operator offers new insights into the existing literature. First, it provides a new interpretation of baseline rationing by defining exclusion thresholds as proportions of the claims \citep[e.g.,][]{Sanchez-Soriano_21_MSS}, while differing from the baseline operators developed by \cite{Hougaard_al_13_SCWE, Hougaard_al_13_AOR} through its dilation mechanism. Second, the resulting paths of awards often align with the Constant-Increasing-Constant (CIC) family of rules, where claimants initially receive nothing, then increasing awards, and finally nothing again once fully compensated \citep[see][]{Thomson_08_SCW}. Our operator produces new members of this family, specifically addressing scenarios where the agent who is allocated the initial gains and losses has the smallest claim. Third, our approach extends the study of claims problems with exogenous weights \citep[e.g.,][]{Hokari_al_03_ET, Thomson_15_ET, Flores-Szwagrzak_15_JET, Harless_17_ET}. By embedding exclusion thresholds directly into the problem, the operator provides a more complex method for constructing weight profiles and modeling asymmetric treatment. Unlike weighted and proportional rules, the operator's modification of the underlying rule arises structurally from the exclusion thresholds, generating a distorted version of the rule that preserves its essence.

The remainder of the paper is organized as follows. Section \ref{sect_model} introduces the standard bilateral claims problem, extends the framework to incorporate exclusion thresholds, and defines the dilation transformation function. This section also shows the application of the dilation transformation to several popular allocation rules. Section \ref{sect_operator} formally defines the exclusion dilation operator. Section \ref{sect_ax_preservation} examines the properties preserved by the operator. Section \ref{sect_ax_charact} provides a formal axiomatic characterization of the operator. Section \ref{sect_remarks} concludes by situating our findings within the existing literature and highlighting the structural differences between the exclusion dilation operator and previously established operators. The Appendix contains the proofs and provides explicit implementations of the operator for several commonly studied allocation rules.


\section{The Model}
\label{sect_model}


\subsection{The benchmark problem}
\label{subsect_claims_problem}

We consider rationing problems in which two agents have claims over a divisible endowment $E\in\mathbb{R}_{+}$. Let $c=(c_{1},c_{2})\in\mathbb{R}^{2}_{+}$ denote the claims profile and let $C=c_{1}+c_{2}\geq E$ denote the aggregate claim. A standard two-agent claims problem is a pair $(c,E)\in\mathbb{R}^{2}_{+}\times\mathbb{R}_{+}$. Let $\mathcal{C}$ denote the collection of all such problems.

An allocation for a problem $(c,E)\in\mathcal{C}$ is a vector $x\in\mathbb{R}^{2}_{+}$ such that $(i)$ $0\leq x_{i} \leq c_{i}$ for each $i=1,2$ (\textit{boundedness}), and $(ii)$ $x_{1}+x_{2}=E$ (\textit{balance}). An allocation rule is a function $R:\mathcal{C}\rightarrow\mathbb{R}^{2}_{+}$ that assigns to each problem $(c,E)\in\mathcal{C}$ an allocation $R(c,E)$.

In the two-agent framework, any continuous and nondecreasing rule can be characterized by the amount allocated to claimant $2$ after assigning a share of the endowment to claimant $1$. More precisely, for any $(c,E)\in\mathcal{C}$, any $x_{1}\in[0,c_{1}]$, and any $R(c,E)$, let $R^{(c,E)}(x_{1})$ be the function that determines claimant $2$'s award when claimant $1$ receives $x_{1}$, so that the balance condition is satisfied. We restrict the analysis to cases in which the solution to this function is unique.\footnote{This requirement is assumed for the sake of clarity in the proofs, but it is not necessary for the results of the paper. The idea behind the assumption is that the path of awards can be characterized solely by the two assignments, regardless of the value of the endowment. The proportional rule satisfies this requirement. The constrained equal awards, constrained equal losses, concede-and-divide, and reverse Talmud rules also satisfy the requirement, provided that claimants $1$ and $2$ are indexed such that $c_{1}\geq c_{2}$ (see subsection \ref{subsect_rules}). In contrast, the constrained egalitarian rule violates the requirement.}


\subsection{Rules for bilateral claims problems}
\label{subsect_rules}

Let us now describe several well-known allocation rules using the function $R^{(c,E)}(x_{1})$. We begin with the proportional rule, which makes awards proportional to claims. \\

\noindent\textbf{\textcolor{MidnightBlue}{Proportional rule:}} For each $(c,E)\in\mathcal{C}$,
  \[
    P(c,E)=\lambda c,
  \]
  where $\lambda\in\mathbb{R}_{+}$ is chosen to satisfy balance. \\

\begin{figure}
  \centering
    \includegraphics[height=44mm]{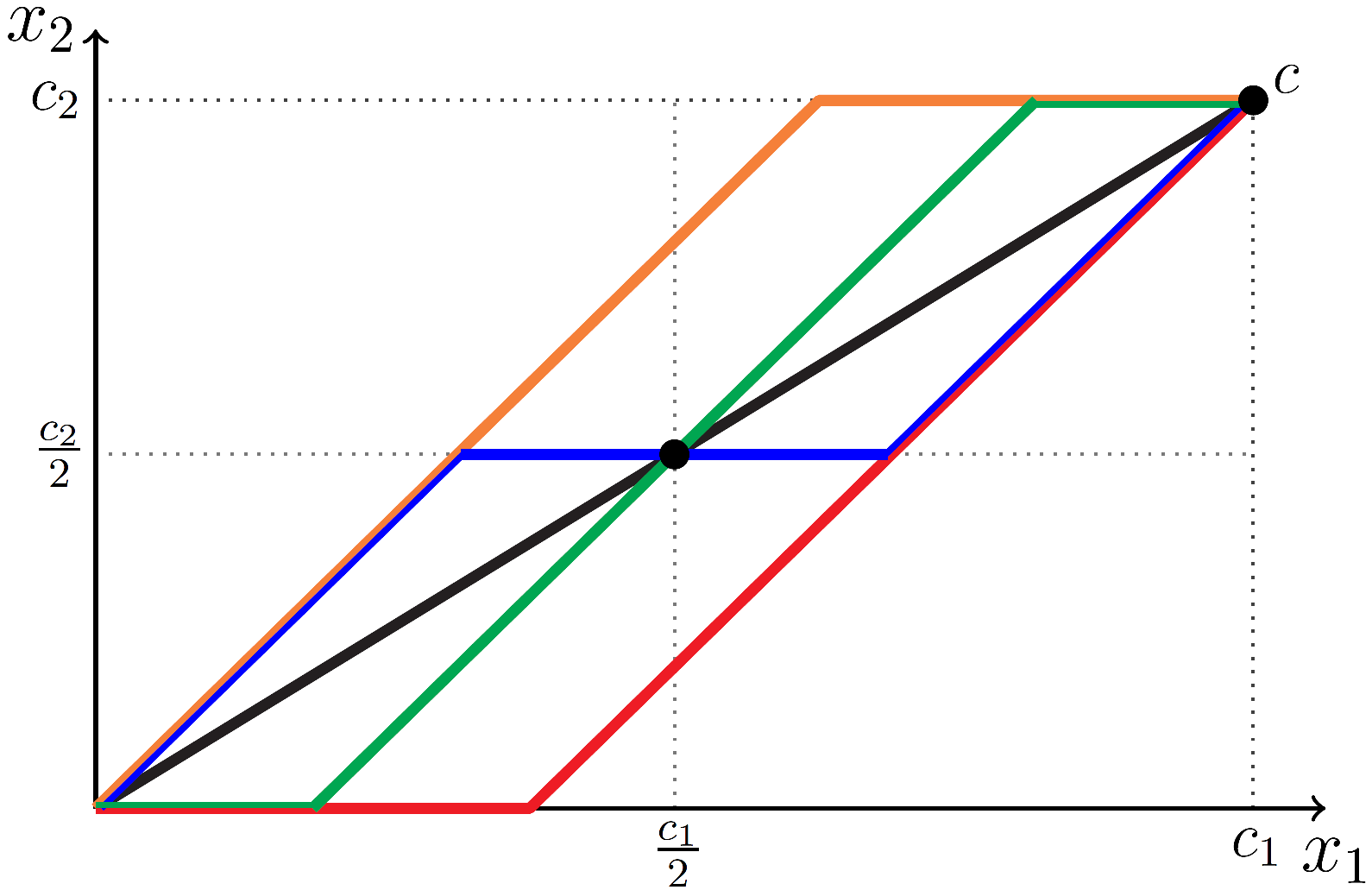}
  \caption{Classic rules for bilateral claims problems}
  \label{fig_classic_rules}
  \end{figure}

The black path in Figure \ref{fig_classic_rules} illustrates this rule for $c=(16,10)$. The proportional rule can also be defined as:
  \[
    P^{(c,E)}(x_{1})=\frac{c_{2}}{c_{1}}x_{1}.
  \] 


Next, the constrained equal awards rule considers equal division in absolute terms, ensuring that no agent receives more than their claim. \\

\noindent\textbf{\textcolor{MidnightBlue}{Constrained equal awards rule:}} For each $(c,E)\in\mathcal{C}$,
  \[
    CEA(c,E)=(\min\{c_{i}, \lambda\})_{i\in N}.
  \]
  where $\lambda\in\mathbb{R}_{+}$ is chosen to satisfy balance. \\

The orange path in Figure \ref{fig_classic_rules} illustrates the constrained equal awards rule. When $c_{1}\geq c_{2}$, it can also be defined as:
  \[
    CEA^{(c,E)}(x_{1})=\min\{x_{1},c_{2}\}.
  \]

 
The dual formula distributes losses equally among claimants, ensuring that none of them receives a negative amount. \\

\noindent\textbf{\textcolor{MidnightBlue}{Constrained equal losses rule:}} For each $(c,E)\in\mathcal{C}$,
  \[
    CEL(c,E)=(\max\{c_{i}-\lambda,0\})_{i\in N}.
  \]
  where $\lambda\in\mathbb{R}_{+}$ is chosen to satisfy balance. \\

The red path in Figure \ref{fig_classic_rules} illustrates the constrained equal losses rule. When $c_{1}\geq c_{2}$, it can also be defined as:
  \[
    CEL^{(c,E)}(x_{1})=\max\{x_{1}-(c_{1}-c_{2}),0\}.
  \]


The next rule only applies when there are two claimants. For each $i=1,2$, concede-and-divide states that by claiming $c_{i}$, agent $i$ concedes to agent $j\neq i$ the amount $E-c_{i}$ if this amount is nonnegative, and 0 otherwise. Symmetrically, agent $j$ concedes $E-c_{j}$  to agent $i$ whenever this amount is nonnegative. The residual is then divided equally between the two agents. \\

\noindent\textbf{\textcolor{MidnightBlue}{Concede-and-divide:}} For $|N|=2$, and for each $(c,E)\in\mathcal{C}$,
  \[
    CD(c,E)=\left(\max\{E-c_{j},0\}+\frac{E-\max\{E-c_{j}\}-\max\{E-c_{i}\}}{2}\right)_{i\in N}.
  \]

The blue path in Figure \ref{fig_classic_rules} illustrates this rule. Note that concede-and-divide applies the constrained equal awards rule when the endowment is less than half of the claims, and the constrained equal losses rule otherwise. When $c_{1}\geq c_{2}$, this rule can also be defined as:
  \[
    CD^{(c,E)}(x_{1})=\min\left\{x_{1}, \max\left\{\frac{c_{2}}{2},x_{1}-(c_{1}-c_{2})\right\}\right\}.
  \]


In the bilateral case, the reverse Talmud rule is the dual of concede-and-divide. It applies the constrained equal losses rule when the endowment is less than half of the claims, and the constrained equal awards rule otherwise. \\

\noindent\textbf{\textcolor{MidnightBlue}{Reverse Talmud rule:}} For each $(c,E)\in\mathcal{C}$ and each $i=1,2$,
  \[
    RT_{i}(c,E)=
    \left \{
      \begin{array}{cl}
        \max\left\{ c_{i}/2-\lambda,0 \right\} \ & \textnormal{if} \ E\leq C/2,
        \\[1.5ex]
         c_{i}/2+\min\{ c_{i}/2,\lambda \} \ & \textnormal{if } \ E\geq C/2,
      \end{array}
    \right.
  \]
  where $\lambda\in\mathbb{R}_{+}$ is chosen to satisfy balance.
  \\

The green path in Figure \ref{fig_classic_rules} illustrates the reverse Talmud rule. When $c_{1}\geq c_{2}$, it can also be defined as:
  \[
    RT^{(c,E)}(x_{1})=\max\left\{ 0, \min\left\{c_{2},x_{1}-\frac{c_{1}-c_{2}}{2}\right\}\right\}.
  \]


\subsection{The extended claims problem with exclusions}
\label{subsect_exclusions}

For any claims problem $(c,E)\in\mathcal{C}$, let $\ell=(\ell_{1},\ell_{2})\in[0,1)^{2}$ and $u=(u_{1},u_{2})\in(0,1]^{2}$ denote the lower and upper exclusions, respectively, defined as proportions of the agents' claims, and satisfying $\ell_{i}<u_{i}$ for each $i=1,2$. Moreover, for some $i\in\{1,2\}$, $\ell_{i}=0$, and for some $j\in\{1,2\}$, $u_{j}=1$. Let $L(c,\ell)=\ell_{1}c_{1}+\ell_{2}c_{2}$ and $U(c,u)=u_{1}c_{1}+u_{2}c_{2}$ be the aggregate lower and upper exclusion thresholds, respectively. Note that $0\leq L(c,\ell)<U(c,u)\leq C$. An extended two-agent claims problem with exclusions consists of a tuple $(c,E,\ell,u)\in\mathbb{R}^{2}_{+}\times\mathbb{R}_{+}\times[0,1)^{2}\times(0,1]^{2}$. Let $\mathcal{S}$ denote the collection of all such problems. For any $(c,E,\ell,u)\in\mathcal{S}$, the exclusion space is the two-dimensional region $(\ell_{1}c_{1},u_{1}c_{1})\times(\ell_{2}c_{2},u_{2}c_{2})$, which is a subspace of the claims space $(0,c_{1})\times (0,c_{2})$.

An allocation for a problem $(c,E,\ell,u)\in\mathcal{S}$ is a vector $x\in\mathbb{R}^{2}_{+}$ such that $(i)$ $0\leq x_{i} \leq c_{i}$ for each $i=1,2$, and $(ii)$ $x_{1}+x_{2}=E$. An extended allocation rule is a function $\delta:\mathcal{S}\rightarrow\mathbb{R}^{2}_{+}$ that assigns to each problem $(c,E,\ell,u)\in\mathcal{S}$ an allocation $\delta(c,E,\ell,u)$.


\subsection{Dilations of rules}
\label{subsect_dilations}

In this paper, we will examine how claimants can be treated asymmetrically based on their exclusion thresholds using an operator. Intuitively, these thresholds limit the range of claims that can effectively participate in the allocation process. We then define our operator by a dilation transformation that proportionally scales any rule from the benchmark problem to the exclusion space.  

The dilation transformation function proportionally shrinks or stretches a function in order to fit it into a different domain. Formally, for any $a,b\in\mathbb{R}_{++}$, any $z\in\mathbb{R}$, and any continuous function $f(z):\mathbb{R}\rightarrow\mathbb{R}$, let us define the dilation transformation of $f(z)$ under $a,b$ as $\widehat{f}(z;a,b):\mathbb{R}\rightarrow\mathbb{R}$ such that $\widehat{f}(z;a,b)=bf(z/a)$. 

Therefore,  for any $(c,E)\in\mathcal{C}$, any $\ell\in[0,1)^{2}$, any $u\in(0,1]^{2}$, and any rule $R(c,E)$, let $\widehat{R}^{(c,E)}(y;s)$ be the dilation transformation of $R(c,E)$ under $s=(s_{1},s_{2})\in(0,1]^{2}$, where $s_{i}=u_{i}-\ell_{i}$ for each $i=1,2$. Note that $y\in[0,s_{1}c_{1}]$ and $\widehat{R}^{(c,E)}(y;s)\in[0,s_{2}c_{2}]$. Thus, the dilated version of the rule has a domain of $(0,s_{1}c_{1})\times(0,s_{2}c_{2})$ and shares resources in the interval $[0,U(c,u)-L(c,\ell)]$.


Let us now present the dilation transformation of the rules introduced in subsection \ref{subsect_rules}. We begin with the proportional rule: \\

\noindent\textbf{\textcolor{MidnightBlue}{Dilated proportional rule under $s$:}} For each $(c,E)\in\mathcal{C}$, each $\ell\in[0,1)^{2}$, and each $u\in(0,1]^{2}$,
  \[
    \widehat{P}^{(c,E)}(y;s)=\frac{s_{2}c_{2}}{s_{1}c_{1}}y.
  \]

\begin{figure}[t!]
  \centering
    \subfloat[The proportional rule]{\label{fig_appli_prop}$
      \includegraphics[height=44mm]{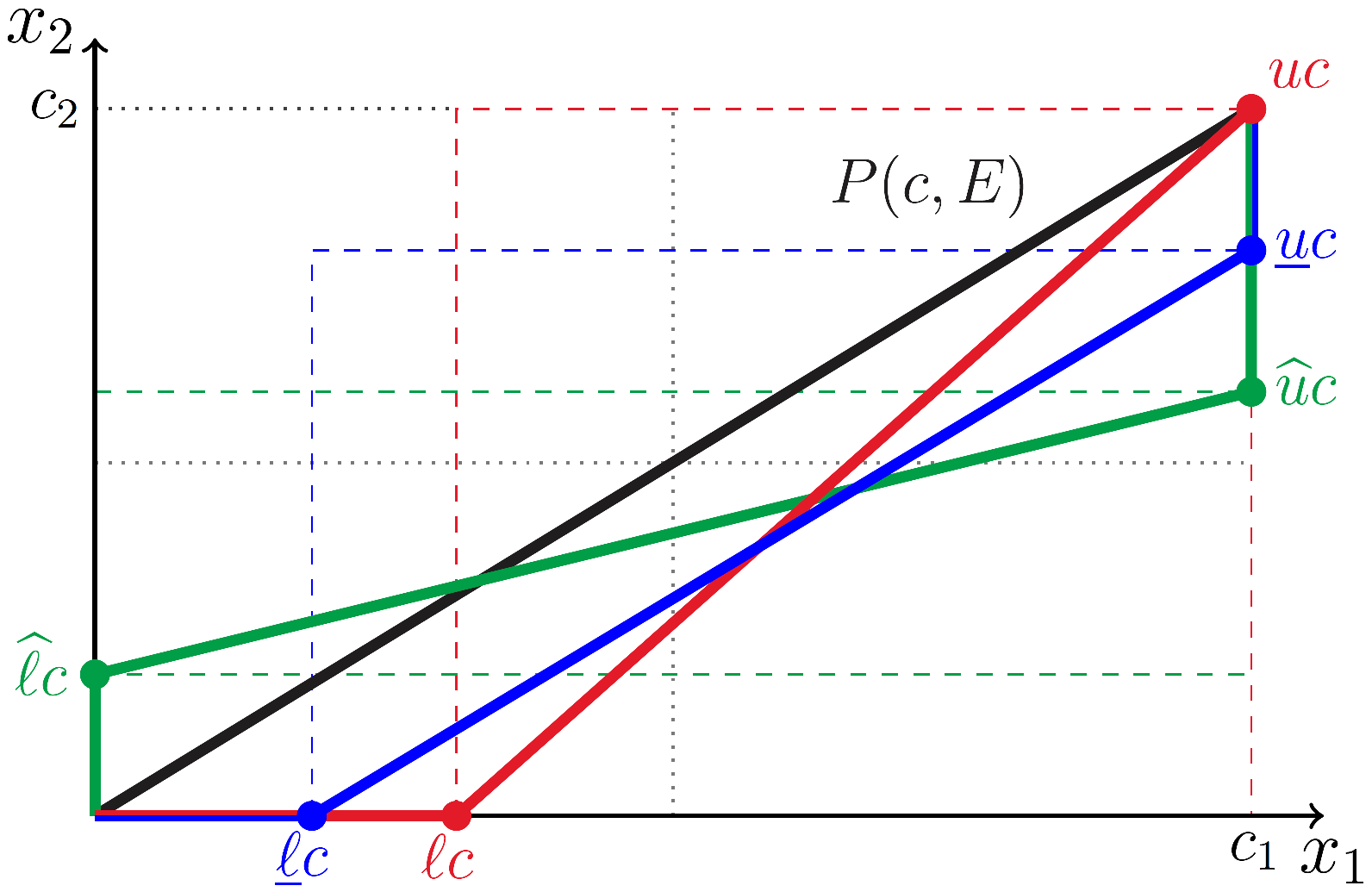}
    $}
      \hspace{0.00cm}
    \subfloat[The constrained equal awards rule]{\label{fig_appli_cea}$
      \includegraphics[height=44mm]{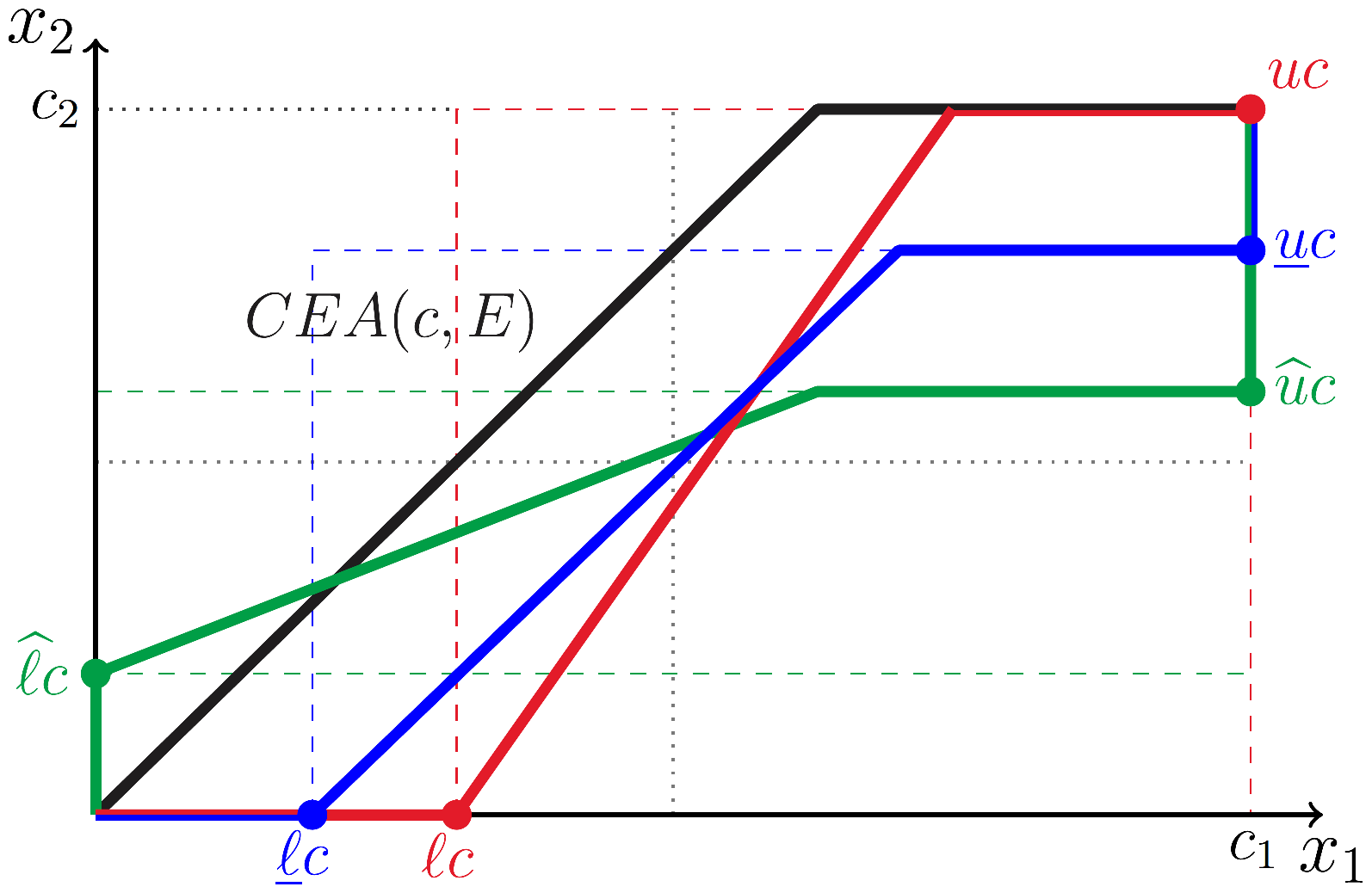}
    $}
      \hspace{0.00cm}
    \subfloat[The constrained equal losses rule]{\label{fig_appli_cel}$
      \includegraphics[height=44mm]{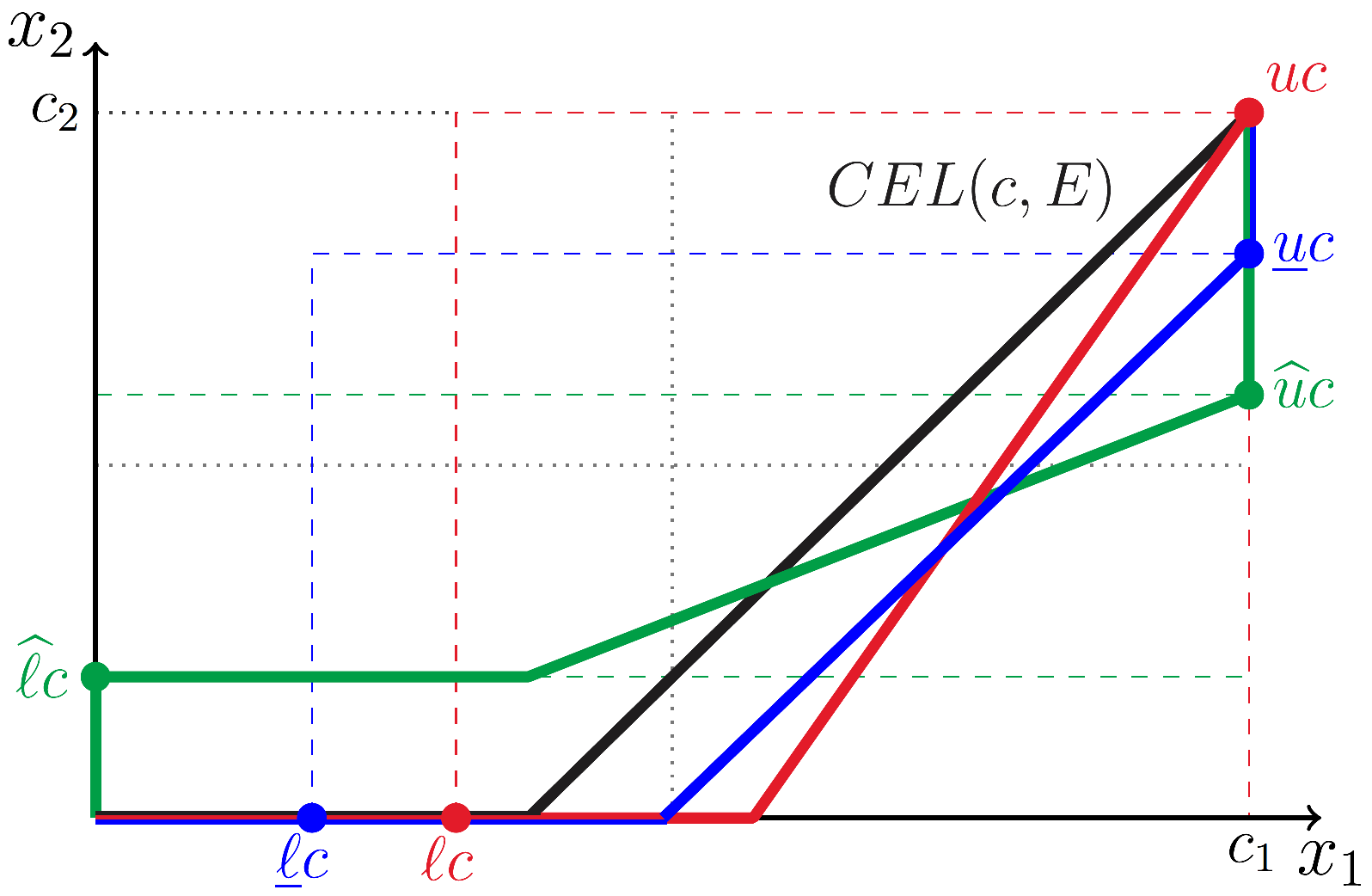}
    $}
      \hspace{0.00cm}
    \subfloat[Concede-and-divide]{\label{fig_appli_cd}$
      \includegraphics[height=44mm]{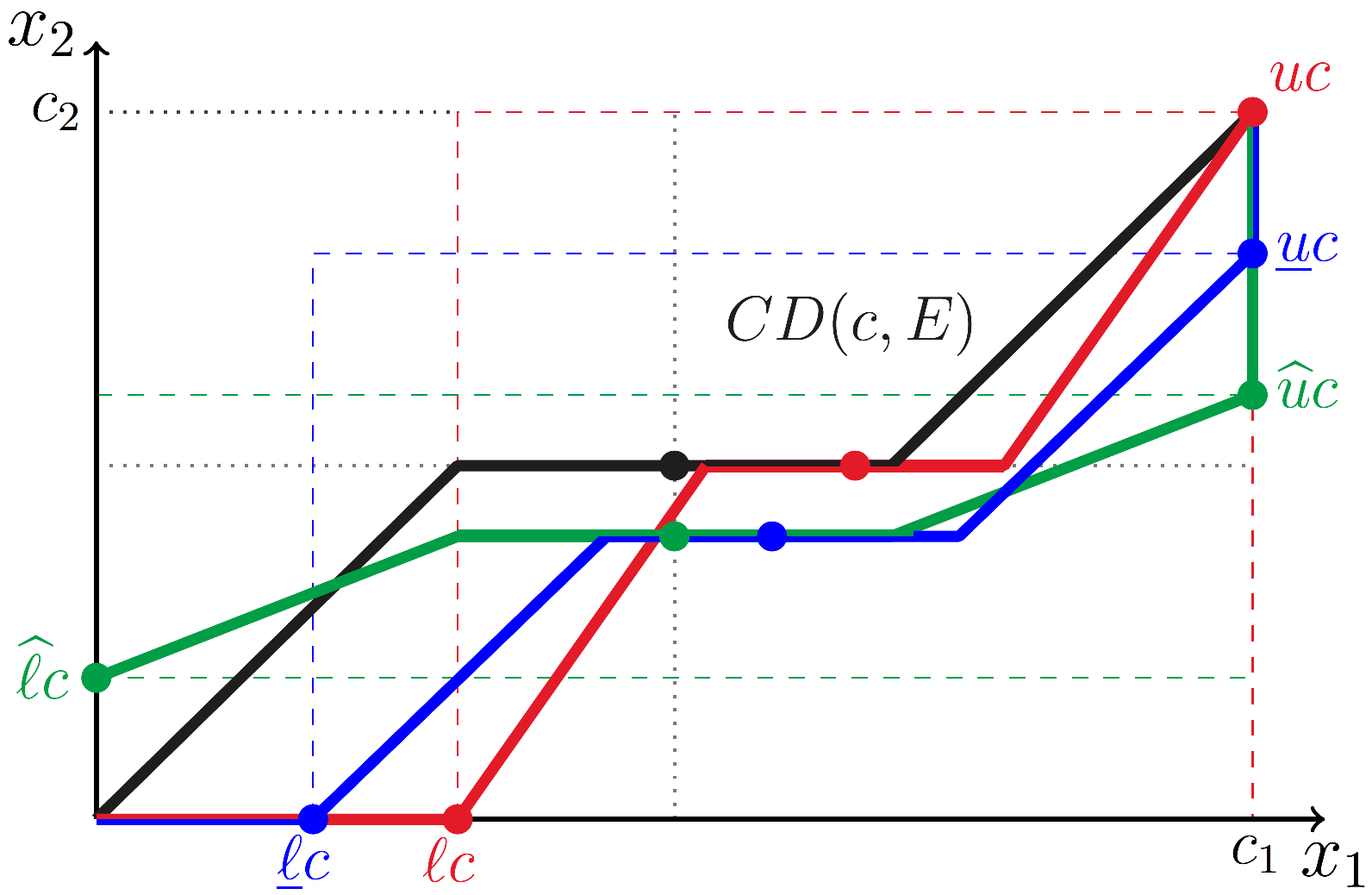}
    $}
      \hspace{0.00cm}
    \subfloat[The reverse Talmud rule]{\label{fig_appli_rt}$
      \includegraphics[height=44mm]{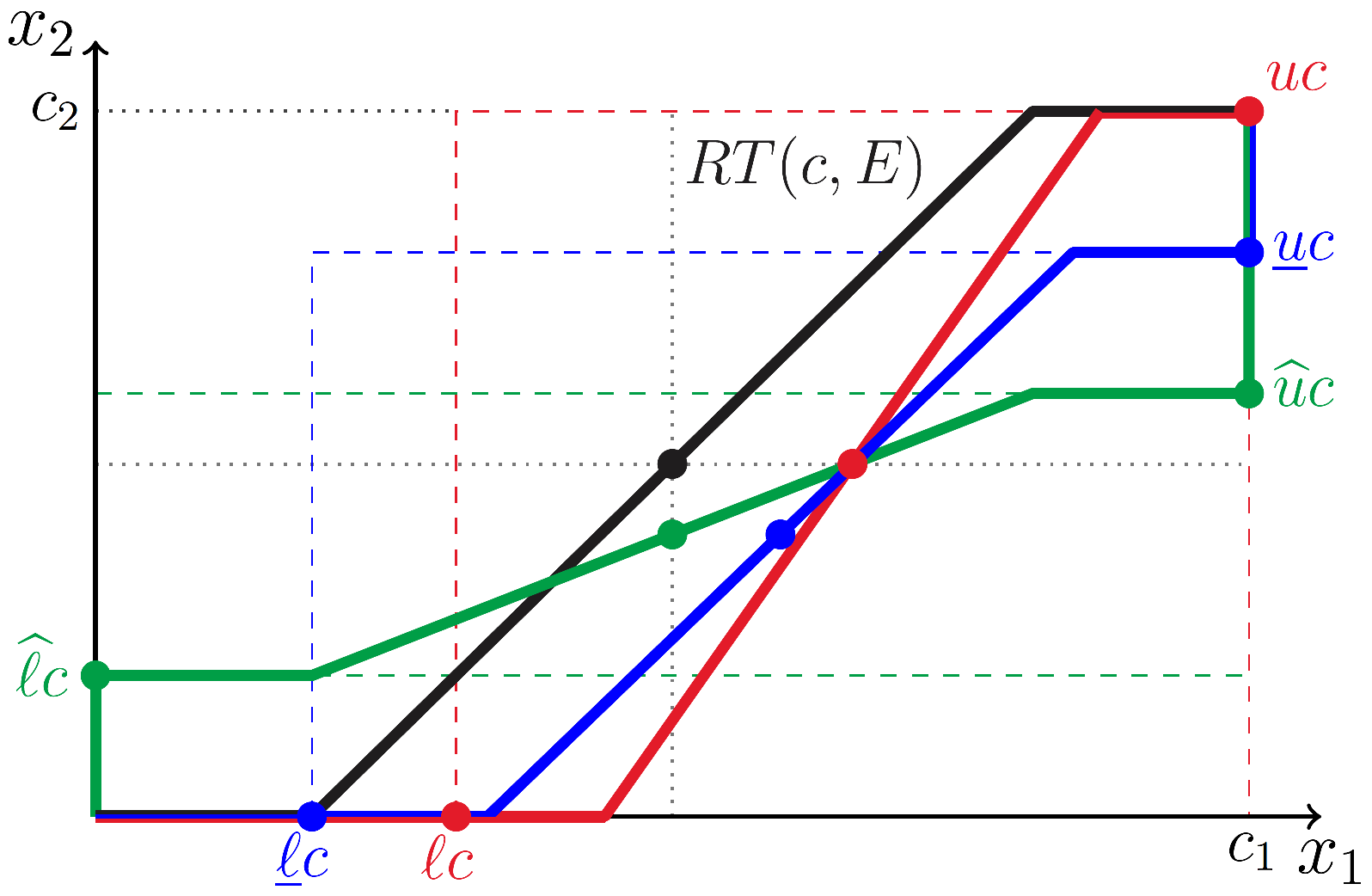}
    $}
    \caption{Dilations and the exclusion dilation operator}
    \label{fig_dilation_rules}
  \end{figure}

Figure \ref{fig_appli_prop} illustrates the dilation transformation of the proportional rule for different exclusion thresholds (the benchmark formula is depicted in black), together with the exclusion spaces related to these thresholds (see the dashed lines). Specifically, we consider the following values: $(\ell,u)=((5/16,0),(1,1))$, $(\underline{\ell},\underline{u})=((3/16,0),(1,0.8))$, and $(\widehat{\ell},\widehat{u})=((0,0.2),(1,0.6))$.\footnote{We represent the exclusion thresholds as $\ell c$ and $uc$, denoting the points $(\ell_{1}c_{1},\ell_{2}c_{2})$ and $(u_{1}c_{1},u_{2}c_{2})$, respectively.} As the picture shows, the dilated version of the rule applies proportionality within the exclusion space $(\ell_{1}c_{1},u_{1}c_{1})\times(\ell_{2}c_{2},u_{2}c_{2})$, which can be defined as a reduced space of size $(0,s_{1}c_{1})\times(0,s_{2}c_{2})$.


Figure \ref{fig_dilation_rules} depicts the dilation transformations of the other rules introduced in subsection 2.2, with the benchmark formulas displayed in black. These transformations are as follows: \\

\noindent\textbf{\textcolor{MidnightBlue}{Dilated constrained equal awards rule under $s$:}} For each $(c,E)\in\mathcal{C}$, each $\ell\in[0,1)^{2}$, and each $u\in(0,1]^{2}$,
  \[
    \widehat{CEA}^{(c,E)}(y;s)=s_{2}\min\left\{ \frac{y}{s_{1}},c_{2}\right\}.
  \]

  
\noindent\textbf{\textcolor{MidnightBlue}{Dilated constrained equal losses rule under $s$:}} For each $(c,E)\in\mathcal{C}$, each $\ell\in[0,1)^{2}$, and each $u\in(0,1]^{2}$,
  \[
    \widehat{CEL}^{(c,E)}(y;s)=s_{2}\max\left\{ \frac{y}{s_{1}}-(c_{1}-c_{2}),0\right\}.
  \]
  
  
\noindent\textbf{\textcolor{MidnightBlue}{Dilated concede-and-divide under $s$:}} For each $(c,E)\in\mathcal{C}$, each $\ell\in[0,1)^{2}$, and each $u\in(0,1]^{2}$,
  \[
    \widehat{CD}^{(c,E)}(y;s)
    =s_{2}\min\left\{\frac{y}{s_{1}}, \max\left\{\frac{c_{2}}{2},\frac{y}{s_{1}}-(c_{1}-c_{2})\right\}\right\}.
  \]


\noindent\textbf{\textcolor{MidnightBlue}{Dilated reverse Talmud rule:}} For each $(c,E)\in\mathcal{C}$, each $\ell\in[0,1)^{2}$, and each $u\in(0,1]^{2}$,
  \[
    \widehat{RT}^{(c,E)}(y;s)
    =s_{2}\max\left\{ 0, \min\left\{c_{2},\frac{y}{s_{1}}-\frac{c_{1}-c_{2}}{2}\right\}\right\}.
  \]


These examples show that the dilation transformation generates a distorted version of the benchmark rule while still retaining its essence. For instance, the constrained equal awards and constrained equal losses rules maintain a combination of one increasing and one constant path of awards, although the slope of the former changes. Concede-and-divide continues to display an increasing-constant-increasing pattern, while the reverse Talmud rule maintains a constant-increasing-constant pattern \citep[see][for a characterization of these types of patterns]{Thomson_08_SCW}. Once again, the dilation transformation modifies the slopes of these two rules.


\section{The exclusion dilation operator}
\label{sect_operator}

We now introduce the exclusion dilation operator. This mechanism is applied to standard rules to adjust the claims and the endowment so that the underlying rule can be applied within the exclusion space. The operator follows a three-stage process. First, it allocates initial gains to claimant $i\in\{1,2\}$ with $\ell_{i}>0$ until the award received by this claimant reaches the lower exclusion threshold $\ell_{i}c_{i}$. Second, within the exclusion space, claimants receive resources according to the dilation transformation of a standard rule. Third, the operator prioritizes claimant $i\in\{1,2\}$ with $u_{i}=1$ in initial losses by fully honoring this agent until the loss of the other claimant reaches the difference between the claim and the upper exclusion threshold, that is, $(1-u_{i})c_{i}$. Formally, \\

\noindent\textbf{\textcolor{Red}{Exclusion dilation operator:}} For each $(c,E,\ell,u)\in\mathcal{S}$, each $R(c,E)$, and each $i=1,2$,
  \[
    \widehat{R}_{i}(c,E,\ell,u)=
    \left \{
      \begin{array}{cl}
        \frac{\ell_{i}c_{i}}{L(c,\ell)}E  \ & \textnormal{if} \ E<L(c,\ell),
        \\[1.25ex]
         \ell_{i}c_{i}+y^{R(c,E)}_{i}(s,c,E-L(c,\ell)) & \textnormal{if} \ E\in[L(c,\ell),U(c,u)],
        \\[1ex]
         u_{i}c_{i}+\frac{(1-u_{i})c_{i}}{C-U(c,u)}(E-U(c,u)) \ \ & \textnormal{if} \ E>U(c,u),
      \end{array}
    \right.
  \]
    where $y^{R(c,E)}_{i}(s,c,E-L(c,\ell))\in[0,s_{i}c_{i}]$ denotes $i$'s award, for any $E-L(c,\ell)\in[0,U(c,u)-L(c,\ell)]$, associated with the dilation transformation of $R(c,E)$ under $s=(s_{1},s_{2})$.
  \\

Note that $y_{i}^{R(c,E)}(s,c,E-L(c,\ell))$, for each $i=1,2$, are the solutions to the system of equations that consists of the balance constraint $y_{1}^{R(c,E)}(s,c,E-L(c,\ell))+y_{2}^{R(c,E)}(s,c,E-L(c,\ell))=E-L(c,\ell)$ and the dilation transformation of the rule $y_{2}^{R(c,E)}(s,c,E-L(c,\ell))=\widehat{R}^{(c,E)}(y_{1}^{R(c,E)}(s,c,E-L(c,\ell));s)$. 

To illustrate how this operator works, let us describe its application to the proportional rule. When solving for the solution of the above-mentioned system of equations, one obtains:
 \[
   y_{i}^{P(c,E)}(s,c,E-L(c,\ell))=\frac{s_{i}c_{i}}{s_{1}c_{1}+s_{2}c_{2}}(E-L(c,\ell)), \ \textnormal{for each} \ i=1,2. 
 \]
Thus, 

  \[
    \widehat{P}_{i}(c,E,\ell,u)=
    \left \{
      \begin{array}{cl}
        \frac{\ell_{i}c_{i}}{L(c,\ell)}E  & \textnormal{if} \ E<L(c,\ell),
        \\[1.5ex]
         \ell_{i}c_{i}+
         \frac{s_{i}c_{i}}{s_{1}c_{1}+s_{2}c_{2}}(E-L(c,\ell)) & \textnormal{if} \ E\in[L(c,\ell),U(c,u)],
        \\[1.5ex]
         u_{i}c_{i}+\frac{(1-u_{i})c_{i}}{C-U(c,u)}(E-U(c,u)) & \textnormal{if} \ E>U(c,u).
      \end{array}
    \right.
  \]
  \\

Figure \ref{fig_appli_prop} depicts the application of the extended proportional rule. According to the exclusion dilation operator, the claimant with a strictly positive lower exclusion threshold receives the initial resources, and the other claimant receives nothing. Once the lower exclusion threshold is reached, the remaining endowment is divided between the two claimants based on the dilation transformation of the proportional rule within the exclusion space. Finally, once the endowment is sufficient to cover the value of the upper exclusion threshold, one claimant stops receiving resources, and the other claimant receives the remaining endowment. In Figure \ref{fig_appli_prop}, the blue and green paths show general applications where $(0,0)< \ell<u<(1,1)$.\footnote{Vector inequalities are denoted $\geq,>,\gg$.} The blue path prioritizes agent $1$ in both gains and losses, while the green path prioritizes agent $2$ in gains and agent $1$ in losses. The red path represents a scenario with no upper exclusion, meaning $u=(1,1)$. 

Figure \ref{fig_dilation_rules} also depicts the application of the exclusion dilation operator to the constrained equal awards, constrained equal losses, concede-and-divide, and reverse Talmud rules. Appendix A provides the explicit application of the operator to these specific rules, as well as to two general nonlinear rules.


\section{Preservation of properties}
\label{sect_ax_preservation}

In this section, we examine which properties are preserved by the exclusion dilation operator. A property is said to be preserved by an operator if, whenever the underlying rule satisfies the property, the extended rule induced by the operator also satisfies it. We consider the following axioms, which are standard in the theory of fair allocation \citep[see][]{Thomson_19_Book}:\footnote{These axioms are formally defined in Appendix B.} \textbf{Equal treatment of equals}: two agents with identical claims should receive equal awards. \textbf{Order preservation}: if agent $i$'s claim is at least as large as agent $j$'s, then agent $i$'s award should be at least as large as agent $j$'s. Likewise, agent $i$'s loss should be at least as large as agent $j$'s. \textbf{Endowment monotonicity}: if the endowment increases, each claimant's award should be at least as large as it was initially. \textbf{Claim monotonicity}: if an agent's claim increases, that agent's award should not decrease. \textbf{Homogeneity}: if the elements of the problem are multiplied by a positive number, all awards should also be multiplied by that same number. \textbf{Midpoint}: if the endowment equals half of the aggregate claims, then each agent should receive half of their claim. \textbf{Self-duality}: the problem of allocating available resources is symmetric to the problem of allocating existing losses. \textbf{Restricted endowment convexity}: keeping the claims vector fixed, if no agent earns nothing or their full claim in two given problems, then the award vector chosen for a convex combination of two endowments is equivalent to the convex combination of the award vectors chosen for each endowment. \textbf{Progressivity}: if agent $i$'s claim is at most as large as agent $j$'s, then agent $i$ should receive proportionally at most as much as agent $j$ does. \textbf{Regressivity}: if agent $i$'s claim is at most as large as agent $j$'s, then agent $i$ should receive proportionally at least as much as agent $j$ does. \textbf{Concavity}: as the endowment increases, successive increments are divided more and more in favor of the smaller claimant. \textbf{Convexity}: as the endowment increases, successive increments are divided more and more in favor of the higher claimant.

Before discussing how the exclusion dilation operator behaves with respect to these axioms, we introduce two specific subdomains of the extended problem. First, exclusions are \textit{symmetric} if, for each agent, the proportion of the claim excluded from gains equals the proportion excluded from losses, i.e., if $\ell_{i}=1-u_{i}$ for all $i=\{1,2\}$. Second, exclusions are \textit{order-preserving} if the agent with the larger claim continues to have the larger claim after subtracting the exclusion thresholds, i.e., if $c_{i}\leq c_{j}$, then $s_{i}c_{i}\leq s_{j}c_{j}$.

The behavior of the operator with respect to the axioms introduced above is summarized in the following theorem. \\

  \begin{theorem}
    The following statements about the exclusion dilation operator hold:
    \begin{enumerate}[i)]
      \item Equal treatment of equals and order preservation are not preserved.
      \item Endowment monotonicity, claim monotonicity, and homogeneity are preserved.
      \item Midpoint and self-duality are preserved if the exclusions are symmetric.
      \item Restricted endowment convexity is preserved within the exclusion space.
      \item Progressivity, regressivity, concavity, and convexity are preserved within the exclusion space if the exclusions are order-preserving.
    \end{enumerate}
    \label{th_preservation}
  \end{theorem}

Naturally, the exclusion dilation operator violates equal treatment and order preservation. The first violation arises when two individuals with identical claims face different exclusion thresholds. The second arises from allowing exclusions for either agent.

Because $s_{1}$ and $s_{2}$ are strictly positive, the dilation transformation does not reverse the nondecreasing pattern of a monotonic rule. Furthermore, proportional changes in the elements of the model induce a secondary proportional dilation of the rule, thereby preserving homogeneity. If one claim changes by a given percentage, the corresponding exclusion thresholds change by the same percentage. Since the paths of awards associated with the original and modified claims cannot intersect, the paths of the dilated versions cannot either, thus satisfying claim monotonicity.

The dilation transformation scales the underlying rule horizontally and vertically by factors $s_{1}$ and $s_{2}$, respectively. If these factors are symmetric, then the transformation produces a symmetric contraction of the rule. This preserves self-duality and the midpoint property within the original space of claims. 

Additionally, a dilation transformation preserves the linearity of the underlying rule. However, once the components of the operator that implement exclusion thresholds are added, the resulting extended rule no longer satisfies restricted endowment convexity. Thus, the operator only preserves this property within the exclusion space.

The exclusion dilation operator does not consider the order of the claims, so it does not generally preserve progressivity, regressivity, concavity, and convexity. However, a dilation transformation produces a proportionally distorted version of the underlying rule. Therefore, the operator preserves these four properties within the exclusion space, as long as the exclusions do not reverse the order of the claims.

Table \ref{tab_preservation of axioms} summarizes how the exclusion dilation operator behaves with respect to the axioms considered in this section.

  \begin{table}[]
  \caption{Preservation of axioms}
  \label{tab_preservation of axioms}
    \begin{tabular*}{\textwidth}{@{\extracolsep\fill}ll}
    \toprule%
      Axioms  &  Preservation \\
    \midrule
        Equal treatment of equals                & No                  \\
        Order preservation                       & No                  \\
        Endowment monotonicity                   & Yes                 \\
        Claim monotonicity                       & Yes                 \\
        Homogeneity                              & Yes                 \\
        Midpoint                                 & $*$                 \\
        Self-duality                             & $*$                 \\
        Restricted endowment convexity           & $\dagger$           \\
        Progressivity / Regressivity             & $\ddagger$          \\
        Concavity / Convexity                    & $\ddagger$          \\
    \bottomrule
    \end{tabular*}
        $*$ \footnotesize With symmetric exclusions. \\
        $\dagger$ \footnotesize Within the exclusion space. \\
        $\ddagger$ \footnotesize Within the exclusion space with order-preserving exclusions.
  \end{table}


\section{The axiomatic characterization}
\label{sect_ax_charact}

We now characterize the exclusion dilation operator. First, we introduce three axioms that describe how exclusion thresholds are handled. Then, we show that the combination of these axioms uniquely determines the operator.

The first and second axioms describe how to allocate initial gains and losses based on exclusion thresholds. Building on the concepts of exclusion and exemption introduced by \cite{Herrero_al_01_MSS}, when the endowment is sufficiently small, one claimant receives nothing; conversely, when the shortage is small relative to claims, one claimant is fully compensated. Whereas those authors identify such claimants based on individual claims, we identify them using the exclusion thresholds. Specifically, when resources are scarce, the individual with the smallest lower exclusion threshold receives nothing, whereas when resources are abundant, the individual with the largest upper exclusion threshold is fully compensated. \\

\noindent\textbf{Full exclusion:} For each $(c,E,\ell,u)\in\mathcal{S}$ and each $i,j\in\{1,2\}$, with $i\neq j$, if $\ell_{i}c_{i}\geq E$, then $\delta_{j}(c,E,\ell,u)=0$.
\\

\noindent\textbf{Null exclusion:} For each $(c,E,\ell,u)\in\mathcal{S}$ and each $i,j\in\{1,2\}$, with $i\neq j$, if $u_{i}c_{i}\leq E-c_{j}$, then $\delta_{j}(c,E,\ell,u)=c_{j}$.
\\

Full exclusion establishes that one claimant should not receive resources until the other claimant has been awarded, at least, their lower exclusion threshold. Null exclusion establishes that one claimant should be fully compensated whenever the remaining endowment is sufficient to cover the other claimant's upper exclusion threshold.

The third axiom determines how resources are allocated once lower and upper exclusion thresholds have been implemented. It establishes a proportional relationship between the solution to the original problem and the solutions to problems defined on reduced claims spaces. Specifically, the allocation within one region of the problem that is equivalent in size to the exclusion space is defined as a proportional scaling of the allocation selected for the full claims problem. Formally, \\

\noindent\textbf{Proportional exclusion invariance:} For each $(c,E)\in\mathcal{C}$, each $\ell\in[0,1)^{2}$, and each $u\in(0,1]^{2}$, there exists $m_{i}\in[0,(1-s_{i})]$, for each $i=1,2$, such that,
  \[
    \begin{array}{l}
        \delta_{i}(c,\widehat{E}+M(c,m),\ell,u)=s_{i}\delta_{i}(c,E,(0,0),(1,1))+m_{i}c_{i},
    \end{array}
  \]
where $m=(m_{1},m_{2})$, $M(c,m)=m_{1}c_{1}+m_{2}c_{2}$, and $\widehat{E}=\sum_{k=1,2}s_{k}\delta_{k}(c,E,(0,0),(1,1))$.
\\

Note that $\sum_{k=1,2}s_{k}\delta_{k}(c,E,(0,0),(1,1))\in[0,s_{1}c_{1}+s_{2}c_{2}]$. Therefore, proportional exclusion invariance establishes the existence of a region of size $(0,s_{1}c_{1})\times(0,s_{2}c_{2})$ in which resources are allocated in proportion to a standard rule.

As the next theorem shows, these three axioms collectively characterize the exclusion dilation operator. \\

  \begin{theorem}
    An extended allocation rule satisfies full exclusion, null exclusion, and proportional exclusion invariance if and only if it is obtained via the exclusion dilation operator.
    \label{th_operator}
  \end{theorem}


\section{Final remarks}
\label{sect_remarks}

We introduced the exclusion dilation operator to extend allocation rules to bilateral claims problems with lower and upper exclusion thresholds. By incorporating these thresholds directly into the structure of the problem, the operator provides a flexible framework for modeling asymmetric treatment in rationing problems. The operator follows a three-stage process. First, lower exclusions are implemented in the distribution of initial gains. Second, resources are allocated within a reduced space by means of a dilation transformation of a standard rule. Finally, upper exclusions are applied in the distribution of initial losses. 

This paper contributes to the literature on fair allocation problems by defining a new class of operator on the space of rules. Previous operators rely on duality, truncation, minimal rights \citep{Thomson_al_08_JET}, weights \citep{Thomson_19_Book}, or baseline adjustments \citep{Hougaard_al_12_JME, Hougaard_al_13_SCWE, Hougaard_al_13_AOR}. In contrast, the exclusion dilation operator uses a dilation transformation based on exclusion thresholds. Unlike approaches that rely on truncation properties or baselines, the operator modifies both the claims and the geometry of the underlying rule. Moreover, unlike weighted and proportional methods, the modification of the rule induced by the operator arises structurally from the exclusion thresholds, generating a distorted version of the underlying rule that preserves its essence.

We presented an axiomatic characterization of the exclusion dilation operator based on full exclusion, null exclusion, and proportional exclusion invariance. Together, these three principles provide a normative foundation for incorporating exclusion thresholds into allocation problems. We also assessed the robustness of the operator by examining which properties of benchmark rules it preserves. While the operator preserves key characteristics of standard rules, such as homogeneity and claim monotonicity, it violates the principles of equal treatment and order preservation. This reflects the intentional asymmetries introduced by exclusion thresholds. Under additional domain restrictions, such as symmetric or order-preserving exclusions, the operator also preserves other properties, including self-duality and progressivity principles. 

Our paper shows how dilation transformations can address asymmetric entitlements or priority structures while preserving the structural coherence of the underlying allocation rule. Beyond its formal contribution, the exclusion dilation operator provides a flexible framework for settings in which fairness principles intersect with legal, fiscal, or policy considerations.


\appendix{}


\section*{Appendix A: Application of the exclusion dilation operator}
\label{app_application_operator}

\begin{figure}[t!]
  \centering
    \subfloat[A strictly nonlinear rule]{\label{fig_appli_convex}$
      \includegraphics[height=44mm]{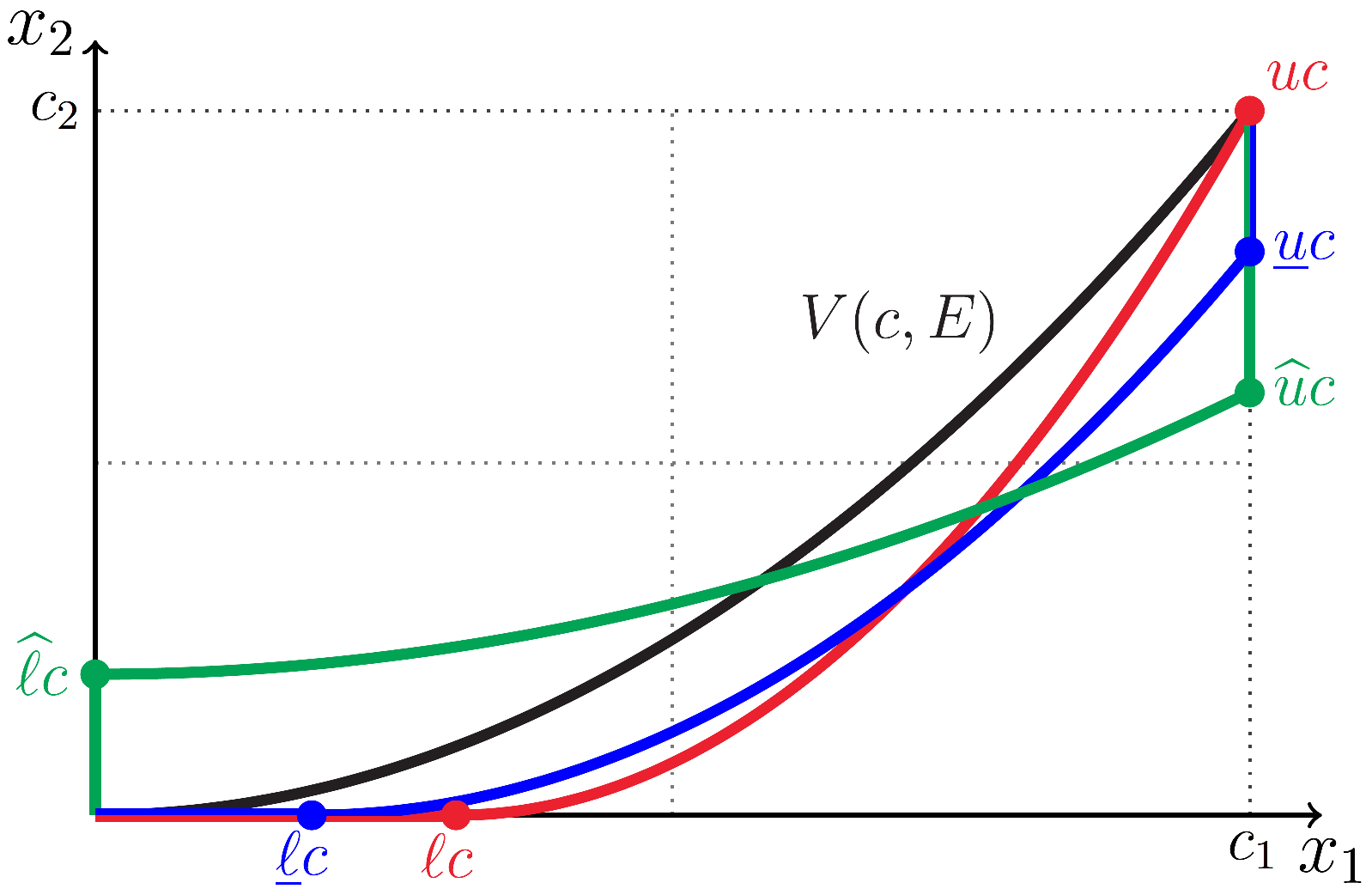}
    $}
      \hspace{0.00cm}
    \subfloat[A differentiable self-dual rule]{\label{fig_appli_sd}$
      \includegraphics[height=44mm]{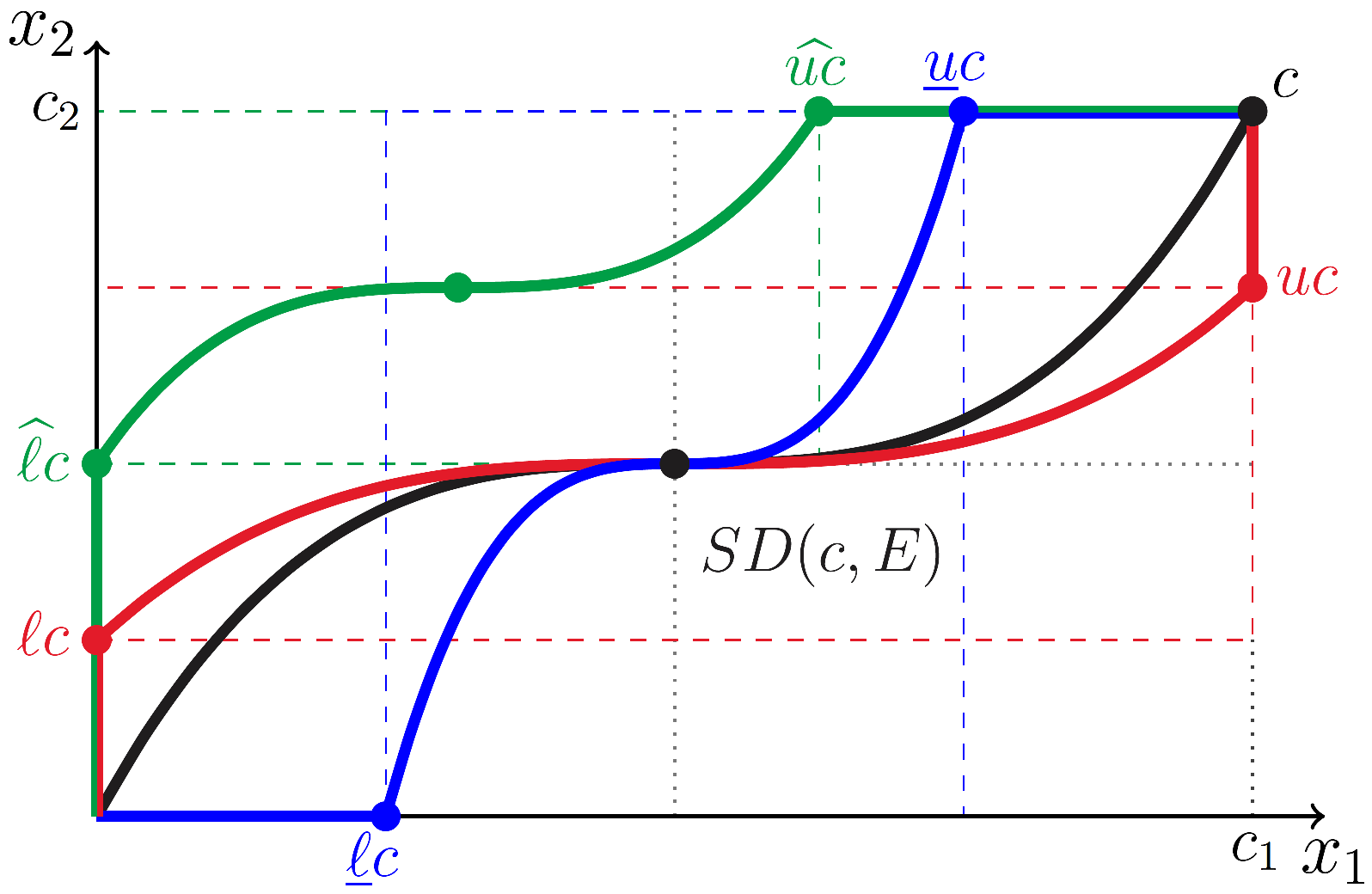}
    $}
    \caption{Application of the exclusion dilation operator}
    \label{fig_application_extra}
  \end{figure}

\begin{enumerate}

\item \noindent\textbf{\textcolor{MidnightBlue}{Constrained equal awards rule:}}
Applying the exclusion dilation operator to this rule yields the following extended rule:
  \[
    \widehat{CEA}_{i}(c,E,\ell,u)=
    \left \{
      \begin{array}{cl}
        \frac{\ell_{i}c_{i}}{L(c,\ell)}E  & \textnormal{if} \ E<L(c,\ell),
        \\[1.5ex]
         \ell_{i}c_{i}+K_{i}(c,E,\ell,u) & \textnormal{if} \ E\in[L(c,\ell),U(c,u)],
        \\[1.5ex]
         u_{i}c_{i}+\frac{(1-u_{i})c_{i}}{C-U(c,u)}(E-U(c,u)) & \textnormal{if} \ E>U(c,u),
      \end{array}
    \right.
  \]
where:
  \[
  \small
    \begin{array}{l}
      K_{1}(c,E,\ell,u)=\max\left\{\frac{s_{1}}{s_{1}+s_{2}}(E-L(c,\ell)),E-L(c,\ell)-s_{2}c_{2} \right\},
 \\[1.5ex]
      K_{2}(c,E,\ell,u)=\min\left\{\frac{s_{2}}{s_{1}+s_{2}}(E-L(c,\ell)),s_{2}c_{2} \right\}.
    \end{array}
  \]
  
Figure \ref{fig_appli_cea} illustrates the application of the extended version of the constrained equal awards rule for $(\ell,u)=((5/16,0),(1,1))$, $(\underline{\ell},\underline{u})=((3/16,0),(1,0.8))$, and $(\widehat{\ell},\widehat{u})=((0,0.2),(1,0.6))$.
   
  
\item \noindent\textbf{\textcolor{MidnightBlue}{Constrained equal losses rule:}}
Applying the exclusion dilation operator to this rule yields the following extended rule:
  \[
    \widehat{CEL}_{i}(c,E,\ell,u)=
    \left \{
      \begin{array}{cl}
        \frac{\ell_{i}c_{i}}{L(c,\ell)}E  & \textnormal{if} \ E<L(c,\ell),
        \\[1.5ex]
         \ell_{i}c_{i}+K_{i}(c,E,\ell,u) & \textnormal{if} \ E\in[L(c,\ell),U(c,u)],
        \\[1.5ex]
         u_{i}c_{i}+\frac{(1-u_{i})c_{i}}{C-U(c,u)}(E-U(c,u)) & \textnormal{if} \ E>U(c,u),
      \end{array}
    \right.
  \]
where:
  \[
  \small
    \begin{array}{l}
      K_{1}(c,E,\ell,u)=\min\left\{\frac{s_{1}}{s_{1}+s_{2}}(E-L(c,\ell)+s_{2}(c_{1}-c_{2})),E-L(c,\ell) \right\},
 \\[1.5ex]
      K_{2}(c,E,\ell,u)=\max\left\{\frac{s_{2}}{s_{1}+s_{2}}(E-L(c,\ell)-s_{1}(c_{1}-c_{2})),0 \right\}.
    \end{array}
  \]
  
Figure \ref{fig_appli_cel} illustrates the application of the extended version of the constrained equal losses rule with the same exclusion thresholds as in the previous example. 
   
  
\item \noindent\textbf{\textcolor{MidnightBlue}{Concede-and-divide:}}
Applying the exclusion dilation operator to this rule yields the following extended rule:
  \[
    \widehat{CD}_{i}(c,E,\ell,u)=
    \left \{
      \begin{array}{cl}
        \frac{\ell_{i}c_{i}}{L(c,\ell)}E  & \textnormal{if} \ E<L(c,\ell),
        \\[1.5ex]
         \ell_{i}c_{i}+K_{i}(c,E,\ell,u) & \textnormal{if} \ E\in[L(c,\ell),U(c,u)],
        \\[1.5ex]
         u_{i}c_{i}+\frac{(1-u_{i})c_{i}}{C-U(c,u)}(E-U(c,u)) & \textnormal{if} \ E>U(c,u),
      \end{array}
    \right.
  \]
where:
  \[
  \small
    \begin{array}{l}
      K_{1}(c,E,\ell,u)
 \\[1.0ex]
      =\max\left\{ \frac{s_{1}(E-L(c,\ell))}{s_{1}+s_{2}}, \min\left\{ E-L(c,\ell)-\frac{s_{2}c_{2}}{2}, \frac{s_{1}}{s_{1}+s_{2}}\left(E-L(c,\ell)+s_{2}(c_{1}-c_{2})\right) \right\} \right\},
 \\[1.5ex]
      K_{2}(c,E,\ell,u)=\min\left\{ \frac{s_{2}(E-L(c,\ell))}{s_{1}+s_{2}},\max\left\{ \frac{s_{2}c_{2}}{2}, \frac{s_{2}}{s_{1}+s_{2}}\left(E-L(c,\ell)-s_{1}(c_{1}-c_{2})\right)\right\} \right\}.
    \end{array}
  \]
  
Figure \ref{fig_appli_cd} illustrates the application of the extended version of concede-and-divide with the same exclusion thresholds as in the previous example.


\item \noindent\textbf{\textcolor{MidnightBlue}{Reverse Talmud rule:}}
Applying the exclusion dilation operator to this rule yields the following extended rule:
  \[
    \widehat{RT}_{i}(c,E,\ell,u)=
    \left \{
      \begin{array}{cl}
        \frac{\ell_{i}c_{i}}{L(c,\ell)}E  & \textnormal{if} \ E<L(c,\ell),
        \\[1.5ex]
         \ell_{i}c_{i}+K_{i}(c,E,\ell,u) & \textnormal{if} \ E\in[L(c,\ell),U(c,u)],
        \\[1.5ex]
         u_{i}c_{i}+\frac{(1-u_{i})c_{i}}{C-U(c,u)}(E-U(c,u)) & \textnormal{if} \ E>U(c,u),
      \end{array}
    \right.
  \]
where:
  \[
  \small
    \begin{array}{l}
      K_{1}(c,E,\ell,u)
 \\[1.0ex]
      =\min\left\{E-L(c,\ell),\max\left\{ E-L(c,\ell)-s_{2}c_{2}, \frac{s_{1}}{s_{1}+s_{2}}\left(E-L(c,\ell)+s_{2}\frac{c_{1}-c_{2}}{2}\right) \right\} \right\},
 \\[1.5ex]
      K_{2}(c,E,\ell,u)=\max\left\{0,\min\left\{s_{2}c_{2}, \frac{s_{2}}{s_{1}+s_{2}}\left(E-L(c,\ell)-s_{1}\frac{c_{1}-c_{2}}{2}\right) \right\} \right\}.
    \end{array}
  \]
  
Figure \ref{fig_appli_rt} illustrates the application of the extended version of the reverse Talmud rule with the same exclusion thresholds as in the previous example.
  
  
\item \noindent\textbf{\textcolor{MidnightBlue}{A strictly nonlinear rule:}} For each $(c,E)\in\mathcal{C}$ and each $x_{1}\in[0,c_{1}]$,
  \[
    V^{(c,E)}(x_{1})=c_{2}\left(\frac{x_{1}}{c_{1}}\right)^2.
  \]  
The dilation transformation of this expression within the exclusion space is given by the following formula:
  \[
    \widehat{V}^{(c,E)}(y;s)=s_{2}c_{2}\left(\frac{y}{s_{1}c_{1}}\right)^2.
  \]
Given the balance constraint, applying the exclusion dilation operator to this formula yields the following extended rule:
  \[
    \widehat{V}_{i}(c,E,\ell,u)=
    \left \{
      \begin{array}{cl}
        \frac{\ell_{i}c_{i}}{L(c,\ell)}E  \ & \textnormal{if} \ E<L(c,\ell),
        \\[1.5ex]
         \ell_{i}c_{i}+K_{i}(T(c,E,\ell,u)) \ & \textnormal{if} \ E\in[L(c,\ell),U(c,u)],
        \\[1.5ex]
         u_{i}c_{i}+\frac{(1-u_{i})c_{i}}{C-U(c,u)}(E-U(c,u)) & \textnormal{if} \ E>U(c,u),
      \end{array}
    \right.
  \]
  where:
  \[
  \small
    \begin{array}{l}
      T(c,E,\ell,u)=\left((s_{1}c_{1})^{2}+4s_{2}c_{2}(E-L(c,\ell))\right)^{1/2},
      \\[1.5ex]
      K_{1}(T(c,E,\ell,u))=\frac{s_{1}c_{1}(T(c,E,\ell,u)-s_{1}c_{1})}{2s_{2}c_{2}},
      \\[1.5ex]
      K_{2}(T(c,E,\ell,u))=E-L(c,\ell)-K_{1}(T(c,E,\ell,u)).
    \end{array}
  \]
  
Figure \ref{fig_appli_convex} illustrates the application of the extended version of this strictly nonlinear rule with the same exclusion thresholds as in the examples represented in Figure \ref{fig_dilation_rules}. The benchmark formula is depicted in black. 
  
   
\item \noindent\textbf{\textcolor{MidnightBlue}{A differentiable self-dual rule:}} For each $(c,E)\in\mathcal{C}$ and each $x_{1}\in[0,c_{1}]$,
  \[
    SD^{(c,E)}(x_{1})=c_{2}\left( \frac{4}{c^{3}_{1}}\left(x_{1}-\frac{c_{1}}{2}\right)^{3}+\frac{1}{2} \right).
  \]
The dilation transformation of this expression within the exclusion space is given by the following formula:
  \[
    \widehat{SD}^{(c,E)}(y;s)
    =s_{2}c_{2}\left( \frac{4}{c^{3}_{1}}\left(\frac{y}{s_{1}}-\frac{c_{1}}{2}\right)^{3}+\frac{1}{2} \right).
  \]
Given the balance constraint, it is possible to obtain the explicit formula for the application of the exclusion dilation operator to this differentiable self-dual rule.\footnote{The explicit formula is too long to be included here, but it can be easily obtained using a software for algebraic calculations.} Figure \ref{fig_appli_sd} illustrates the application of the extended version of this rule for different exclusion thresholds (the benchmark formula is depicted in black). In this particular example $c=(16,10)$, $(\ell,u)=((0,0.25),(1,0.75))$, $(\underline{\ell},\underline{u})=((0.25,0),(0.75,1))$, and $(\widehat{\ell},\widehat{u})=((0,0.5),(10/16,1))$. 

\end{enumerate}


\section*{Appendix B: The inventory of properties of division rules}
\label{app_invent_rules}

\noindent\textcolor{RedOrange}{\textbf{Equal treatment of equals:}} For each $(c,E)\in\mathcal{C}$ and each pair $\{i,j\}\in\{1,2\}$, if $c_{i}=c_{j}$, then $R_{i}(c,E)=R_{j}(c,E)$.
\\

\noindent\textcolor{RedOrange}{\textbf{Order preservation:}} For each $(c,E)\in\mathcal{C}$ and each pair $\{i,j\}\in\{1,2\}$, if $c_{i}\leq c_{j}$, then $R_{i}(c,E)\leq R_{j}(c,E)$. Moreover, $c_{i}-R_{i}(c,E)\leq c_{j}-R_{j}(c,E)$.
\\

\noindent\textcolor{RedOrange}{\textbf{Endowment monotonicity:}} For each $(c,E)\in\mathcal{C}$ and each $E'>E$ such that $E'\leq C$, one has $R_{i}(c,E')\geq R_{i}(c,E)$ for each $i=1,2$.
\\

\noindent\textcolor{RedOrange}{\textbf{Claim monotonicity:}} For each $(c,E)\in\mathcal{C}$, each $i=1,2$, and each $c'_{i}>c_{i}$, one has $R_{i}((c'_{i},c_{j}),E)\geq R_{i}(c,E)$.
\\

\noindent\textcolor{RedOrange}{\textbf{Homogeneity:}} For each $(c,E)\in\mathcal{C}$ and each $\lambda>0$, one has $R(\lambda c,\lambda E)=\lambda R(c,E)$.
\\

\noindent\textcolor{RedOrange}{\textbf{Midpoint:}} For each $(c,E)\in\mathcal{C}$, if $E=C/2$, then $R(c,E)=c/2$.
\\

\noindent\textcolor{RedOrange}{\textbf{Self-duality:}} For each $(c,E)\in\mathcal{C}$, one has $R(c,E)=c-R(c,C-E)$.
\\

\noindent\textcolor{RedOrange}{\textbf{Restricted endowment convexity:}} For each $(c,E)\in\mathcal{C}$, $E'\in\mathbb{R}_{++}$ such that $C\geq\max\{E,E'\}$, and each $\lambda\in[0,1]$, if $0\ll R(c,E),R(c,E') \ll c$, then $R(c,\lambda E+(1-\lambda)E')=\lambda R(c,E)+(1-\lambda)R(c,E')$.
\\

\noindent\textcolor{RedOrange}{\textbf{Progressivity:}} For each $(c,E)\in\mathcal{C}$ and each pair $\{i,j\}\in\{1,2\}$ with $i\neq j$, if $0<c_{i}\leq c_{j}$, then $R_{i}(c,E)/c_{i}\leq R_{j}(c,E)/c_{j}$.
\\

\noindent\textcolor{RedOrange}{\textbf{Regressivity:}} For each $(c,E)\in\mathcal{C}$ and each pair $\{i,j\}\in\{1,2\}$ with $i\neq j$, if $0<c_{i}\leq c_{j}$, then $R_{i}(c,E)/c_{i}\geq R_{j}(c,E)/c_{j}$.
\\

\noindent\textcolor{RedOrange}{\textbf{Concavity:}} For each $(c,E)\in\mathcal{C}$, $\{E',E''\}\in\mathbb{R}_{+}$ such that $0<E<E'<E''\leq C$, and each pair $\{i,j\}\in\{1,2\}$, if $0<c_{i}\leq c_{j}$, then
\[
\frac{R_{j}(c,E')-R_{j}(c,E)}{R_{i}(c,E')-R_{i}(c,E)}\geq\frac{R_{j}(c,E'')-R_{j}(c,E')}{R_{i}(c,E'')-R_{i}(c,E')}.
\]

\noindent\textcolor{RedOrange}{\textbf{Convexity:}} For each $(c,E)\in\mathcal{C}$, $\{E',E''\}\in\mathbb{R}_{+}$ such that $0<E<E'<E''\leq C$, each pair $\{i,j\}\in\{1,2\}$, if $0<c_{i}\leq c_{j}$, then
\[
\frac{R_{j}(c,E')-R_{j}(c,E)}{R_{i}(c,E')-R_{i}(c,E)}\leq\frac{R_{j}(c,E'')-R_{j}(c,E')}{R_{i}(c,E'')-R_{i}(c,E')}.
\]


\section*{Appendix C: Proofs}


\subsection*{Proof of Theorem \ref{th_preservation}}
\label{proof_th_preservation}

\begin{figure}
  \centering
    \subfloat[Equal treatment of equals and order preservation]{\label{fig_proof_th_presv_ete}$
      \includegraphics[height=44mm]{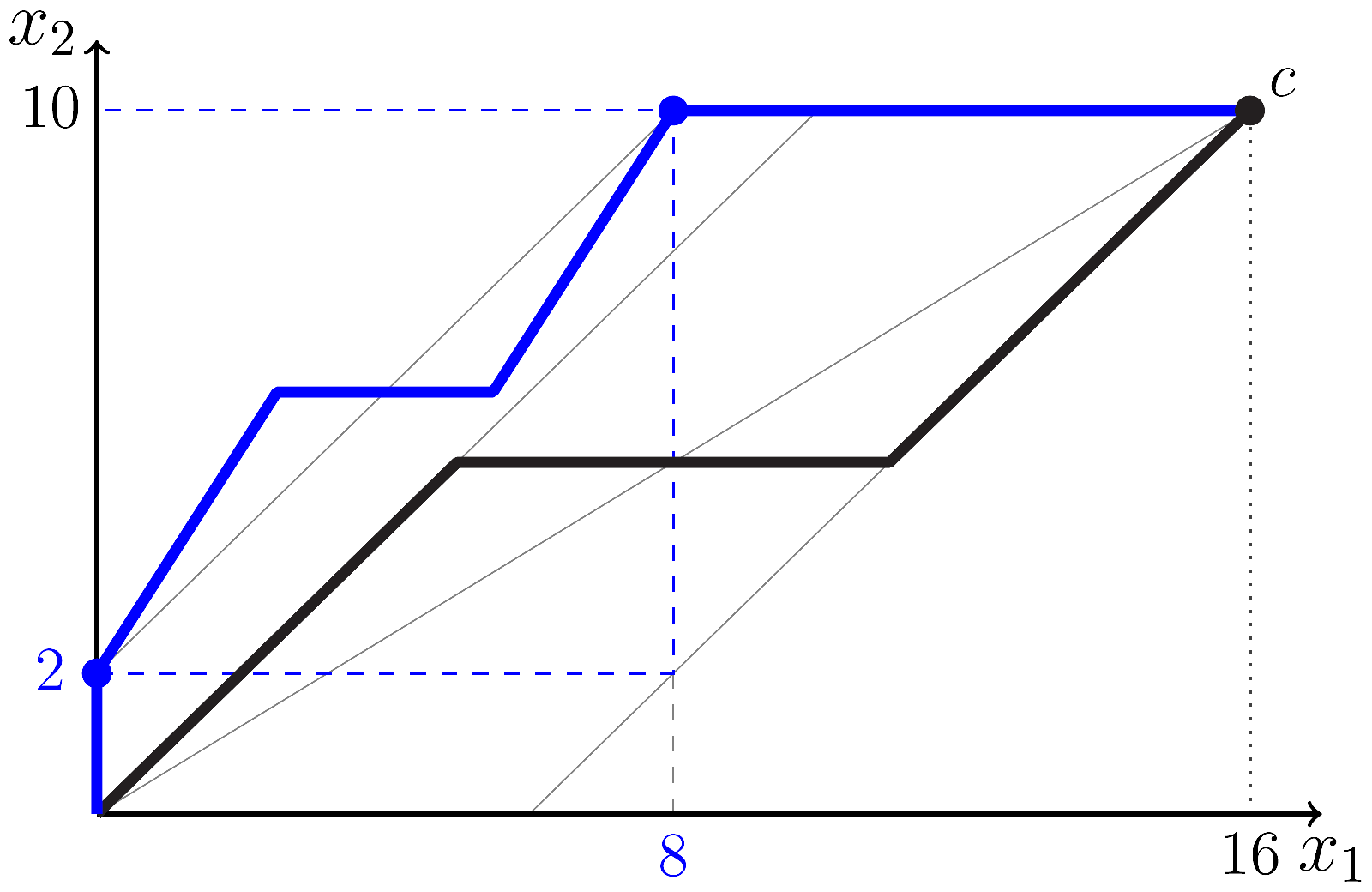}
    $}
      \hspace{0.00cm}
    \subfloat[Restricted endowment convexity]{\label{fig_proof_th_rec_op}$
      \includegraphics[height=44mm]{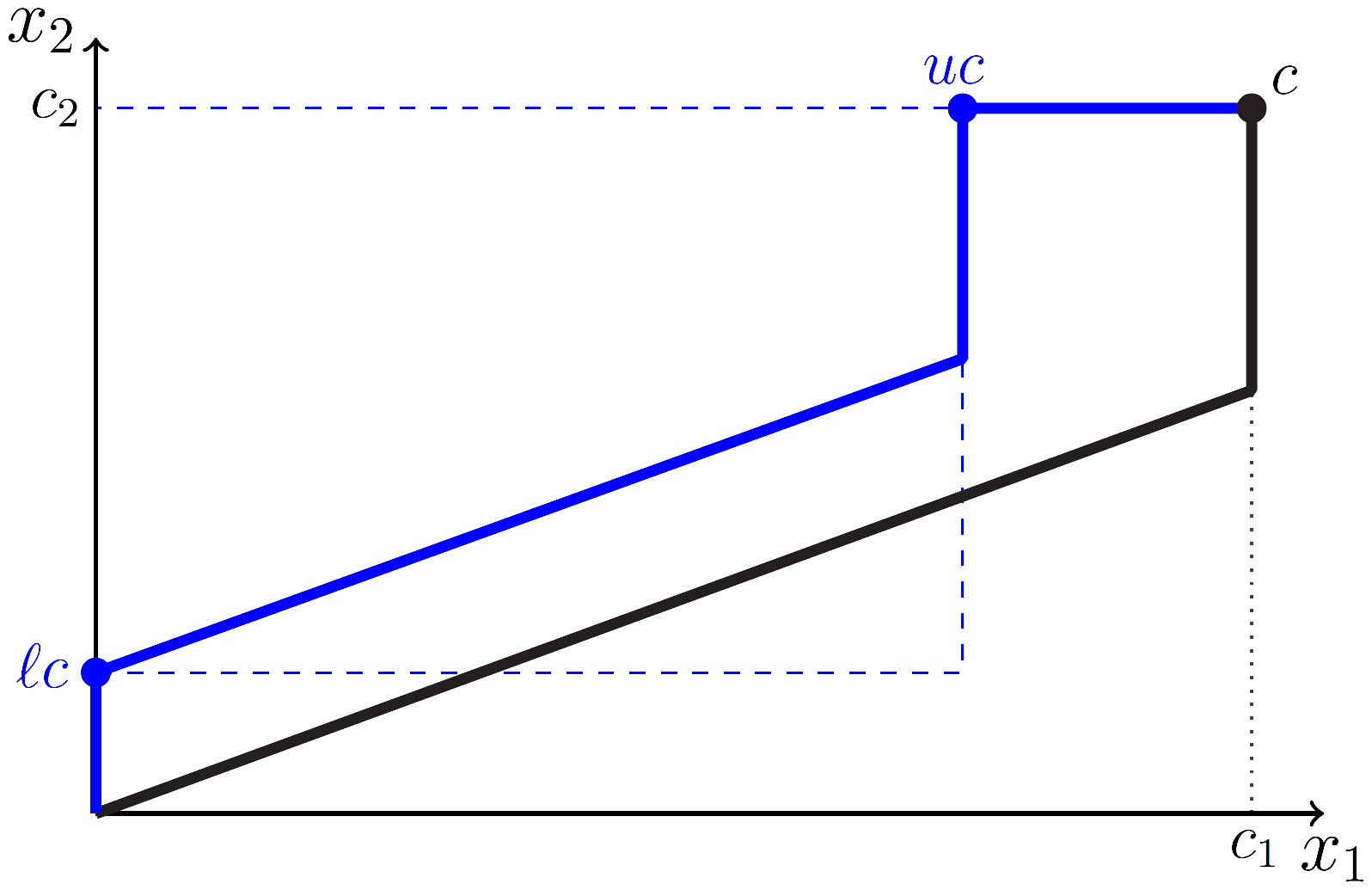}
    $}
    \caption{Properties not preserved by the exclusion dilation operator}
    \label{fig_proof_th_presv}
  \end{figure}

  \begin{enumerate}[$\cdot$]
    \item \textbf{Equal Treatment of Equals}: Figure \ref{fig_proof_th_presv_ete} shows that this property is not preserved when applied to concede-and-divide. It is also not preserved within the exclusion space. For example, in the blue region, the two individuals have identical claims after deducting the exclusion levels, but the dilated version of the rule does not treat them equally.

    \item \textbf{Order Preservation}: Since this axiom implies equal treatment of equals, and based on the previous result, order preservation is not preserved under the exclusion dilation operator (see Figure \ref{fig_proof_th_presv_ete}).

    \item \textbf{Endowment Monotonicity}: Consider $(\ell,u)\in[0,1)^{2}\times(0,1]^{2}$ and a rule $R(c,E)$ characterized by a nondecreasing path of awards $R^{(c,E)}(x_{1})$, with $x_{1}\in[0,c_{1}]$. For any $E,E'\in\mathbb{R}_{+}$, if $E<E'\leq L(c,\ell)$ or $E'>E\geq U(c,u)$, the new allocation is such that one of the two claimants receives all the extra resources and the other receives none. If $ L(c,\ell)\leq E<E'\leq U(c,u)$, the extra endowment is allocated according to the dilation transformation of $R(c,E)$ within the exclusion space, that is, by $\widehat{R}^{(c,E)}(y;s)$, with $y\in[0,s_{1}c_{1}]$. With the vertical dilation (the effect of claimant $2$'s exclusions in the rule), the path of awards in the exclusion space is characterized by $s_{2}R^{(c,E)}(y)$. Since the original rule is nondecreasing, and $s_{2}\in(0,1]$, this path is also nondecreasing. With the horizontal dilation (the effect of claimant $1$'s exclusions in the rule), the path of awards in the exclusion space is characterized by $R^{(c,E)}(y/s_{1})$, which, since $s_{1}\in(0,1]$, is also nondecreasing. Finally, if $E< L(c,\ell)<E'$ or $E< U(c,u)<E'$, then by the continuity of the operator, no one receives less as the endowment increases.

    \item \textbf{Claim Monotonicity}: Consider $c'_{i}>c_{i}$ for some agent $i\in\{1,2\}$, and let $R(c,E)$ be a rule that satisfies claim monotonicity. For the two-claimant problem, and with a continuous rule, claim monotonicity is equivalent to ensuring that the paths of awards for two claims vectors differing by one coordinate do not cross, though they may touch \citep[see][]{Thomson_19_Book}. Consider the case where $\ell_{i}=0$, which implies $L(c,\ell)=L(c',\ell)=\ell_{j}c_{j}$, and hence $\widehat{R}_{i}((c'_{i},c_{j}),E,\ell,u)=\widehat{R}_{i}(c,E,\ell,u)=0$, for any $E \leq L(c,\ell)$. Assume $E\in[L(c,\ell),U(c,u)]$. Since the rule is claim monotonic, then, for any $x_{j}\leq c_{j}$ and $E'>E$, one has $R_{i}(c,E)\leq R_{i}((c'_{i},c_{j}),E')$, where $E=R_{i}(c,E)+x_{j}$ and $E'=R_{i}(c',E')+x_{j}$. This implies $s_{i}R_{i}(c,E)\leq s_{i}R_{i}(c',E')$, that is, within the exclusion space the paths of awards for the two claims vectors do not cross. Therefore, $\widehat{R}_{i}((c'_{i},c_{j}),E,\ell,u)\geq\widehat{R}_{i}(c,E,\ell,u)$. Consider now $E\geq U(c,u)$. Due to the previous result, $\widehat{R}_{i}((c'_{i},c_{j}),U(c,u),\ell,u)\geq\widehat{R}_{i}(c,U(c,u),\ell,u)$. If $u_{j}=1$, then $\widehat{R}_{j}(c,E,\ell,u)=c_{j}$ and $\widehat{R}_{i}(c,E,\ell,u)=E-c_{j}\in[u_{i}c_{i},c_{i}]$. This implies $\widehat{R}_{j}(c',E,\ell,u)\leq c_{j}$, and therefore claim monotonicity is preserved. If $u_{j}<1$, implying $u_{i}=1$, then $\widehat{R}_{i}(c,E,\ell,u)=c_{i}$ and $\widehat{R}_{j}(c,E,\ell,u)=E-c_{i}\in[u_{j}c_{j},c_{j}]$. Since $u_{i}c_{i}=c_{i}<c'_{i}=u_{i}c'_{i}$, and the fact that the paths do not cross, then $\widehat{R}_{j}(c',E,\ell,u)\leq u_{j}c_{j}$, unless $\widehat{R}_{i}(c',E,\ell,u)=c'_{i}$, and therefore claim monotonicity is preserved. Consider now the case where $\ell_{i}>0$, which implies that $\ell_{i}c'_{i}>\ell_{i}c_{i}>0$. For any $E \leq L(c,\ell)$, one has $\widehat{R}_{i}((c'_{i},c_{j}),E,\ell,u)=\widehat{R}_{i}(c,E,\ell,u)=E$. If $C\leq L(c',\ell)$, then claim monotonicity is preserved. Let us consider $C>L(c',\ell)$, and any $E\geq L(c,\ell)$. If $C\leq U(c',u)=u_{i}c'_{i}+u_{j}c_{j}$, then, since the paths of awards for the two claims vectors do not cross within the exclusion space, and since $\ell_{i}c'_{i}>\ell_{i}c_{i}$, one has $\widehat{R}_{i}((c'_{i},c_{j}),E,\ell,u)\geq \widehat{R}_{i}(c,E,\ell,u)$. If $C>U(c',u)$ and $E\leq U(c,u)$, then the result follows, once again, from the fact that the paths do not cross. Finally, consider $E>U(c,u)$. If $u_{j}=1$, one has $\widehat{R}_{j}(c,E,\ell,u)=c_{j}$, and therefore $\widehat{R}_{j}(c',E,\ell,u)\leq c_{j}$. If $u_{j}<1$, implying $u_{i}=1$, one has $\widehat{R}_{i}(c,E,\ell,u)=c_{i}$ and $\widehat{R}_{j}(c,E,\ell,u)=E-c_{i}\in[u_{j}c_{j},c_{j}]$. Since $u_{i}c_{i}=c_{i}<c'_{i}=u_{i}c'_{i}$, then $\widehat{R}_{j}(c',E,\ell,u)\leq u_{j}c_{j}$, unless $\widehat{R}_{i}(c',E,\ell,u)=c'_{i}$, and therefore claim monotonicity is preserved.

    \item \textbf{Homogeneity}: Let $R(c,E)$ be a homogeneous rule characterized by the pair of awards $(x_{1},R^{(c,E)}(x_{1}))$. For any $E \leq L(c,\ell)$ and $i=1,2$, one has $\widehat{R}_{i}(c,E,\ell,u)=\frac{\ell_{i}c_{i}}{L(c,\ell)}E$. Then, for any $\lambda>0$, one has $\lambda E \leq \lambda L(c,\ell)$ and $\widehat{R}_{i}(\lambda c,\lambda E,\ell,u)=\frac{\ell_{i}\lambda c_{i}}{\lambda L(c,\ell)}\lambda E=\lambda\frac{\ell_{i}c_{i}}{L(c,\ell)}E$. For any $E \geq U(c,u)$ and $i=1,2$, one has $\widehat{R}_{i}(c,E,\ell,u)=u_{i}c_{i}+\frac{c_{i}(1-u_{i})}{C-U(c,u)}(E-U(c,u))$. Then, for any $\lambda>0$, one has $\lambda E \geq \lambda U(c,u)$ and $\widehat{R}_{i}(\lambda c,\lambda E,\ell,u)=u_{i}\lambda c_{i}+\frac{\lambda c_{i}(1-u_{i})}{\lambda C-\lambda U(c,u)}(\lambda E-\lambda U(c,u))=\lambda(u_{i}c_{i}+\frac{c_{i}(1-u_{i})}{C-U(c,u)}(E-U(c,u))$. Let us then consider $E\in[L(c,\ell),U(c,u)]$. The path of awards of the extended rule is $\widehat{R}(c,E,\ell,u)=(\ell_{1}c_{1}+y_{1},\ell_{2}c_{2}+y_{2})$, where $(y_{1},y_{2})$ are the solutions to the system $y_{1}+y_{2}=E-L(c,\ell)$ and $y_{2}=s_{2}R^{(c,E)}(y_{1}/s_{1})$. Therefore, $y_{1}=E-L(c,\ell)-s_{2}R^{(c,E)}(y_{1}/s_{1})$. Since $R(c,E)$ is homogeneous, then $\widehat{R}(\lambda c,\lambda E,\ell,u)=(\ell_{1}\lambda c_{1}+y'_{1},\ell_{2}\lambda c_{2}+s_{2}\lambda R^{(c,E)}(y_{1}/s_{1}))$, where $y'_{1}=\lambda(E-L(c,\ell)-s_{2}R^{(c,E)}(y_{1}/s_{1}))=\lambda y_{1}$.
    
    \item \textbf{Midpoint}: Implementing the exclusion dilation operator does not preserve the midpoint property for the proportional rule when $\ell_{1}>0$ and $u_{2}<1$. Therefore, let us focus on the case of symmetric exclusions, and let $R(c,E)$ be a rule that satisfies $R(c,C/2)=c/2$. The value of the original midpoint in the dilated version of the rule is given by $(s_{1}c_{1}/2,s_{2}c_{2}/2)$. If the exclusions are symmetric, then $u_{i}=1-\ell_{i}$ for all $i=1,2$. Based on the assumptions of the model, this implies either a vertical dilation with $s_{1}=1$ and $s_{2}=1-2\ell_{2}\leq 1$ (see the red path in figure \ref{fig_appli_sd}), or a horizontal dilation with $s_{1}=1-2\ell_{1}\leq 1$ and $s_{2}=1$ (see the blue path in figure \ref{fig_appli_sd}). This also implies that $L(c,\ell)<C/2<U(c,u)$. In either scenario, the path of awards in the exclusion space passes through the awards $s_{i}c_{i}/2$ and $c_{j}/2$, where $i\in\{1,2\}$ is the individual with $\ell_{i}\geq0$, and $j\neq i$ is the individual with $\ell_{j}=0$. The endowment associated with these awards is $(1-2\ell_{i})c_{i}/2+c_{j}/2=C/2-\ell_{i}c_{i}$. Adding the value of the lower exclusion threshold yields $\widehat{R}_{k}(c,C/2,\ell,u)=c_{k}/2$, for each $k\in\{i,j\}$.
    
    \item \textbf{Self-duality}: Implementing the exclusion dilation operator does not preserve self-duality for the proportional rule when $\ell_{1}>0$ and $u_{2}<1$. Therefore, let us focus on the case of symmetric exclusions, and let $R(c,E)$ be a rule that satisfies $R(c,E)=c-R(c,C-E)$. This rule can be characterized by the pair of awards $(x_{1},R^{(c,E)}(x_{1}))$. If the exclusions are symmetric, then $u_{i}=1-\ell_{i}$ for all $i=1,2$. Based on the assumptions of the model, this implies either a vertical dilation with $s_{1}=1$ and $s_{2}=1-2\ell_{2}\leq 1$, or a horizontal dilation with $s_{1}=1-2\ell_{1}\leq 1$ and $s_{2}=1$. For any $E\leq L(c,\ell)$ or $E\geq U(c,u)$, the result is a direct consequence of the symmetry of the exclusions. Let us then consider $E\in[L(c,\ell),U(c,u)]$. We start with the vertical dilation (see the red path in figure \ref{fig_appli_sd}). The path of awards is $\widehat{R}(c,E,\ell,u)=(y_{1},\ell_{2}c_{2}+y_{2})$, where $(y_{1},y_{2})$ are the solutions to the system $y_{1}+y_{2}=E-L(c,\ell)$ and $y_{2}=(1-2\ell_{2})R^{(c,E)}(y_{1})$, with $y_{1}\in[0,c_{1}]$ and $y_{2}\in[0,(1-2\ell_{2})c_{2}]$. Likewise, $\widehat{R}(c,C-E,\ell,u)=(y'_{1},\ell_{2}c_{2}+y'_{2})$, where $(y'_{1},y'_{2})$ are the solutions to the system $y'_{1}+y'_{2}=C-E-L(c,\ell)$ and $y'_{2}=(1-2\ell_{2})R^{(c,E)}(y'_{1})$, with $y'_{1}\in[0,c_{1}]$ and $y'_{2}\in[0,(1-2\ell_{2})c_{2}]$. Note that $y'_{1}+y'_{2}=C-E-L(c,\ell)=c_{1}+s_{2}c_{2}+2\ell_{2}c_{2}-E-\ell_{2}c_{2}=c_{1}+s_{2}c_{2}-(E-L(c,\ell))$. That is, $(y_{1},y_{2})$ and $(y'_{1},y'_{2})$ are the solutions, within the exclusion space, to sharing gains $E-L(c,\ell)$ and losses $E-L(c,\ell)$, respectively. Consequently, $y_{1}+s_{2}R^{(c,E)}(y_{1})=c_{1}+s_{2}c_{2}-y'_{1}-s_{2}R^{(c,E)}(y'_{1})$. By self-duality, which implies $R^{(c,E)}(y_{1})=c_{2}-R^{(c,E)}(c_{1}-y_{1})$, the equality can be rewritten as $s_{2}\left(R^{(c,E)}(y'_{1})-R^{(c,E)}(c_{1}-y_{1})\right)=(c_{1}-y_{1})-y'_{1}$. Since the path of awards is continuos and nondecreasing, and $s_{2}>0$, the previous equality only holds, for every rule $R(c,E)$, when $c_{1}-y_{1}=y'_{1}$. One has then $c_{2}-\widehat{R}_{2}(c,E,\ell,u)=c_{2}-(\ell_{2}c_{2}+(1-2\ell_{2})R^{(c,E)}(y_{1}))$, and by the self-duality of $R(c,E)$, $c_{2}-(\ell_{2}c_{2}+(1-2\ell_{2})(c_{2}-R^{(c,E)}(y'_{1})))=\ell_{2}c_{2} +(1-2\ell_{2})R^{(c,E)}(y'_{1})=\widehat{R}_{2}(c,C-E,\ell,u)$. Now, we consider the case of a horizontal dilation (see the blue path in figure \ref{fig_appli_sd}). The path of awards is $\widehat{R}(c,E,\ell,u)=(\ell_{1}c_{1}+y_{1},y_{2})$, where $(y_{1},y_{2})$ are the solutions to the system $y_{1}+y_{2}=E-L(c,\ell)$ and $y_{2}=R^{(c,E)}(y_{1}/(1-2\ell_{1}))$, with $y_{1}\in[0,(1-2\ell_{1})c_{1}]$ and $y_{2}\in[0,c_{2}]$. Likewise, $\widehat{R}(c,C-E,\ell,u)=(\ell_{1}c_{1}+y'_{1},y'_{2})$, where $(y'_{1},y'_{2})$ are the solutions to the system $y'_{1}+y'_{2}=C-E-L(c,\ell)$ and $y'_{2}=R^{(c,E)}(y'_{1}/(1-2\ell_{1}))$, with $y'_{1}\in[0,(1-2\ell_{1})c_{1}]$ and $y'_{2}\in[0,c_{2}]$. Once again, $(y_{1},y_{2})$ and $(y'_{1},y'_{2})$ are the solutions, within the exclusion space, to sharing gains $E-L(c,\ell)$ and losses $E-L(c,\ell)$, respectively. Consequently, $y_{1}+R^{(c,E)}(y_{1}/s_{1})=s_{1}c_{1}+c_{2}-y'_{1}-R^{(c,E)}(y'_{1}/s_{1})$. By self-duality, which implies $R^{(c,E)}(y_{1}/s_{1})=c_{2}-R^{(c,E)}(c_{1}-y_{1}/s_{1})$, the equality can be rewritten as $\left(R^{(c,E)}(y'_{1}/s_{1})-R^{(c,E)}(c_{1}-y_{1}/s_{1})\right)/s_{1}=(c_{1}-y_{1}/s_{1})-y'_{1}/s_{1}$. Since the path of awards is continuos and nondecreasing, and $s_{1}>0$, the previous equality only holds, for every rule $R(c,E)$, when $c_{1}-y_{1}/s_{1}=y'_{1}/s_{1}$. One has then $c_{2}-\widehat{R}_{2}(c,E,\ell,u)=c_{2}-R^{(c,E)}(y_{1}/(1-2\ell_{1}))$, and by the self-duality of $R(c,E)$, $c_{2}-(c_{2}-R^{(c,E)}(y'_{1}/(1-2\ell_{1})))=R^{(c,E)}(y'_{1}/(1-2\ell_{1}))=\widehat{R}_{2}(c,C-E,\ell,u)$. Due to balance, one has $E-y_{2}=\ell_{1}c_{1}+y_{1}$. Moreover, $c_{1}-(\ell_{1}c_{1}+y_{1})=c_{1}-(\ell_{1}c_{1}+(1-2\ell_{1})c_{1}-y'_{1})=\ell_{1}c_{1}+y'_{1}$. Finally, by the continuity of the operator, for any symmetric dilation transformation of a rule that satisfies self-duality one has $\widehat{R}(c,E,\ell,u)=c-\widehat{R}(c,C-E,\ell,u)$.

    \item \textbf{Restricted Endowment Convexity}: Implementing the exclusion dilation operator does not preserve restricted endowment convexity in the claims space (see Figure \ref{fig_proof_th_rec_op}). Therefore, let us focus on the exclusion space, and let $R(c,E)$ be a standard rule that satisfies restricted endowment convexity. Under this property, the points of the rule that satisfy $0<R(c,E)<c$ can be characterized as $R^{(c,E)}(x_{1})=k+\alpha x_{1}$, with $k\in\mathbb{R}$ and $\alpha\in\mathbb{R}_{+}$. The dilation transformation of this expression is $\widehat{R}^{(c,E)}(y;s)=s_{2}k+s_{2}\alpha y/s_{1}$. Since $s_{2}/s_{1}>0$, this function characterizes a linear relationship in the interior region the exclusion space.

    \item \textbf{Progressivity / Regressivity}: Consider $0<c_{2}\leq c_{1}$. If $\ell_{2}>0$ (respectively, $\ell_{1}>0$), then the path of awards associated with the exclusion dilation operator lies above (respectively, below) the proportional distribution when resources are small, which violates progressivity (respectively, regressivity) in the claims space. Let us then focus on the exclusion space. The dilation transformation of any point $(R_{1},R_{2})$ belonging to $R(c,E)$ into this space is given by $(s_{1}R_{1},s_{2}R_{2})$. If the exclusions are order-preserving, then claimant $1$ still has the highest claim, i.e. $0<s_{2}c_{2}\leq s_{1}c_{1}$. If the rule satisfies progressivity, then $R_{1}/c_{1}\geq R_{2}/c_{2}$, and therefore $s_{1}R_{1}/(s_{1}c_{1})\geq s_{2}R_{2}/(s_{2}c_{2})$ (note that $s_{1},s_{2}>0$). Consequently, within the exclusion space claimant $1$ receives proportionally at least as much as claimant $2$ does. The proof for regressivity is analogous.

    \item \textbf{Concavity / Convexity}: Consider $0<c_{2}\leq c_{1}$ and a concave rule $R(c,E)$. If $\ell_{2}>0$, then for any $0<E<E'\leq\ell_{2}c_{2}$ one has $\frac{0-0}{E'-E}=0$, which violates concavity in the claims space. Let us then focus on the exclusion space. The dilation transformation of any point $(R_{1},R_{2})$ belonging to $R(c,E)$ into this space is given by $(s_{1}R_{1},s_{2}R_{2})$. If the exclusions are order-preserving, then claimant $1$ still has the highest claim, i.e. $0<s_{2}c_{2}\leq s_{1}c_{1}$. According to concavity, for any $0<E<E'<E''\leq C$ one has $\frac{R_{1}(c,E')-R_{1}(c,E)}{R_{2}(c,E')-R_{2}(c,E)}\geq \frac{R_{1}(c,E'')-R_{1}(c,E')}{R_{2}(c,E'')-R_{2}(c,E')}$. When applying the dilation transformation to these points, one has $\frac{s_{1}(R_{1}(c,E')-R_{1}(c,E))}{s_{2}(R_{2}(c,E')-R_{2}(c,E))}\geq \frac{s_{1}(R_{1}(c,E'')-R_{1}(c,E'))}{s_{2}(R_{2}(c,E'')-R_{2}(c,E'))}$. Consequently, concavity is preserved within the exclusion space (note that $s_{1},s_{2}>0$). Now, let us consider a convex rule $R(c,E)$. If $\ell_{1}>0$, then for any $0<E<E'\leq\ell_{1}c_{1}$ one has $\frac{E'-E}{0-0}\rightarrow+\infty$, which violates convexity in the claims space. The rest of the proof is analogous to that for concavity.
    
  \end{enumerate}

\subsection*{Proof of Theorem \ref{th_operator}}
\label{proof_th_operator}

It is easy to show that the exclusion dilation operator satisfies \textit{full exclusion} and \textit{null exclusion}. It also satisfies \textit{proportional exclusion invariance}. For each $(c,E)\in\mathcal{C}$, each $\ell\in[0,1)^{2}$, each $u\in(0,1]^{2}$, any standard rule $R(c,E)$, and each $i=1,2$, let $x_{i}=\widehat{R}_{i}(c,E,(0,0),(1,1))$ and $y_{i}=\widehat{R}_{i}(c,\widehat{E}+L(c,\ell),\ell,u)-\ell_{i}c_{i}$, where $\widehat{E}=s_{1}x_{1}+s_{2}x_{2}$. Therefore, $y_{2}-s_{2}x_{2}=s_{1}x_{1}-y_{1}$. By the application of the dilation transformation of the operator, $x_{2}=R^{(c,E)}(x_{1})$ and $y_{2}=\widehat{R}^{(c,E)}(y_{1};s)=s_{2}R^{(c,E)}(y_{1}/s_{1})$. Consequently, $s_{2}\left(R^{(c,E)}(y_{1}/s_{1})-R^{(c,E)}(x_{1})\right)=s_{1}\left(x_{1}-y_{1}/s_{1}\right)$. Since the path of awards is continuos and nondecreasing, and $s_{1},s_{2}>0$, the previous equality only holds, for every rule $R(c,E)$, when $x_{1}=y_{1}/s_{1}$. Moreover, $y_{2}=s_{1}x_{1}+s_{2}x_{2}-y_{1}$, which implies $y_{2}/s_{2}=x_{2}+s_{2}/s_{1}(x_{1}-y_{1}/s_{1})$. Because of the previous result, $x_{2}=y_{2}/s_{2}$. Since $y_{i}=s_{i}x_{i}$ for each $i=1,2$, and by assuming $m=\ell$, one has then $\widehat{R}_{i}(c,\widehat{E}+M(c,m),\ell,u)=m_{i}c_{i}+s_{i}R_{i}(c,E)$, where $\widehat{E}=\sum_{k=1,2}\widehat{R}_{k}(c,E,(0,0),(1,1))$. \\

Now, we focus on the converse implication of Theorem \ref{th_operator}. For any $(c,E,\ell,u)\in\mathcal{S}$, let $\delta$ be an extended rule satisfying \textit{full exclusion}, \textit{null exclusion}, and \textit{proportional exclusion invariance}. Without loss of generality, we assume $0=\ell_{1}<u_{1}<1$ and $0<\ell_{2}<u_{2}=1$. Let $L(c,\ell)=\ell_{2}c_{2}$ and $U(c,u)=u_{1}c_{1}+c_{2}$, which imply $L(c,\ell)<U(c,u)$.  \\

\noindent \textbf{Case 1}, $E<L(c,\ell)$: \\
By \textit{full exclusion} $\delta_{1}(c,E,\ell,u)=0$, and hence $\delta_{2}(c,E,\ell,u)=E$. Note that for any $R(c,E)$ one has $\widehat{R}_{1}(c,E,\ell,u)=0c_{1}/(0c_{1}+\ell_{2}c_{2})E=0$ and $\widehat{R}_{1}(c,E,\ell,u)=\ell_{2}c_{2}/(0c_{1}+\ell_{2}c_{2})E=E$. \\

\noindent \textbf{Case 2}, $E>U(c,u)$: \\
By \textit{null exclusion}, $\delta_{2}(c,E,\ell,u)=c_{2}$, and hence $\delta_{1}(c,E,\ell,u)=E-c_{2}$. Note that for any $R(c,E)$ one has $\widehat{R}_{1}(c,E,\ell,u)=u_{1}c_{1}+(1-u_{1})c_{1}/(c_{1}+c_{2}-u_{1}c_{1}-c_{2})(E-u_{1}c_{1}-c_{2})=E-c_{2}$, and $\widehat{R}_{2}(c,E,\ell,u)=c_{2}+(1-1)c_{2}/(c_{1}+c_{2}-u_{1}c_{1}-c_{2})(E-u_{1}c_{1}-c_{2})=c_{2}$. \\

\noindent\textbf{Case 3, $E\in[L(c,\ell),U(c,u)]$}: \\
By the previous results $\delta(c,L(c,\ell),\ell,u)=(0,\ell_{2}c_{2})$ and  $\delta(c,U(c,u),\ell,u)=(u_{1}c_{1},c_{2})$. The region between these two allocations is $(0,u_{1}c_{1})\times(\ell_{2}c_{2},c_{2})$, which is equal in size to $(0,s_{1}c_{1})\times(0,s_{2}c_{2})$. Then, due to \textit{proportional exclusion invariance}, one has $m=\ell$. Since $\delta_{i}(c,0,(0,0),(1,1))=0$ for each $i=1,2$, by \textit{proportional exclusion invariance} $\delta_{i}(c,L(c,\ell),\ell,u)=s_{i}\delta_{i}(c,0,(0,0),(1,1))+\ell_{i}c_{i}=\ell_{i}c_{i}$. Likewise, $\delta_{i}(c,C,(0,0),(1,1))=c_{i}$, and hence, $\delta_{i}(c,U(c,u),\ell,u)=s_{i}\delta_{i}(c,C,(0,0),(1,1))+\ell_{i}c_{i}=s_{i}c_{i}+\ell_{i}c_{i}=u_{i}c_{i}$, where $U(c,u)=s_{1}c_{1}+s_{2}c_{2}+L(c,\ell)$. Moreover, by \textit{proportional exclusion invariance}, for any $E\in[0,C]$, one also has $\delta_{i}(c,\widehat{E}+L(c),\ell,u)=s_{i}\delta_{i}(c,E,(0,0),(1,1))+\ell_{i}c_{i}$, where $\widehat{E}=\sum_{k=1,2}s_{k}\delta_{k}(c,E,(0,0),(1,1))\in[0,s_{1}c_{1}+s_{2}c_{2}]$. This implies, after subtracting the aggregate lower exclusion thresholds, a dilation transformation of all the points belonging to $\delta_{i}(c,E,(0,0),(1,1))$ within the region $(\ell_{1}c_{1},u_{1}c_{1})\times(\ell_{2}c_{2},u_{2}c_{2})$, whose size is equal to $(0,s_{1}c_{1})\times(0,s_{2}c_{2})$. \\

To summarize:
  \[
    \delta_{i}(c,E,\ell,u)=
    \left \{
      \begin{array}{cl}
        \frac{\ell_{i}c_{i}}{L(c,\ell)}E  & \textnormal{if} \ E<L(c,\ell),
        \\[1.25ex]
         \ell_{i}c_{i}+y_{i}^{\delta(c,E)}(s,c,E-L(c,\ell)) & \textnormal{if} \ E\in[L(c,\ell),U(c,u)],
        \\[1ex]
         u_{i}c_{i}+\frac{(1-u_{i})c_{i}}{C-U(c,u)}(E-U(c,u)) & \textnormal{if} \ E>U(c,u).
      \end{array}
    \right.
  \]
  \\
  where $y_{i}^{\delta(c,E)}(s,c,E-L(c,\ell))$ is the dilation transformation of claimant $i$'s award according to $\delta(c,E,(0,0),(1,1))$ within the space $(0,s_{1}c_{1})\times(0,s_{2}c_{2})$. \\

Let $R$ be a function such that for any $(c,E)\in\mathcal{C}$,
\[
  R(c,E)=\delta(c,E,(0,0),(1,1)).
\]

This implies that $\delta(c,E,\ell,u)=\widehat{R}(c,E,\ell,u)$. \\ 

\begin{figure}
  \centering
    \includegraphics[height=44mm]{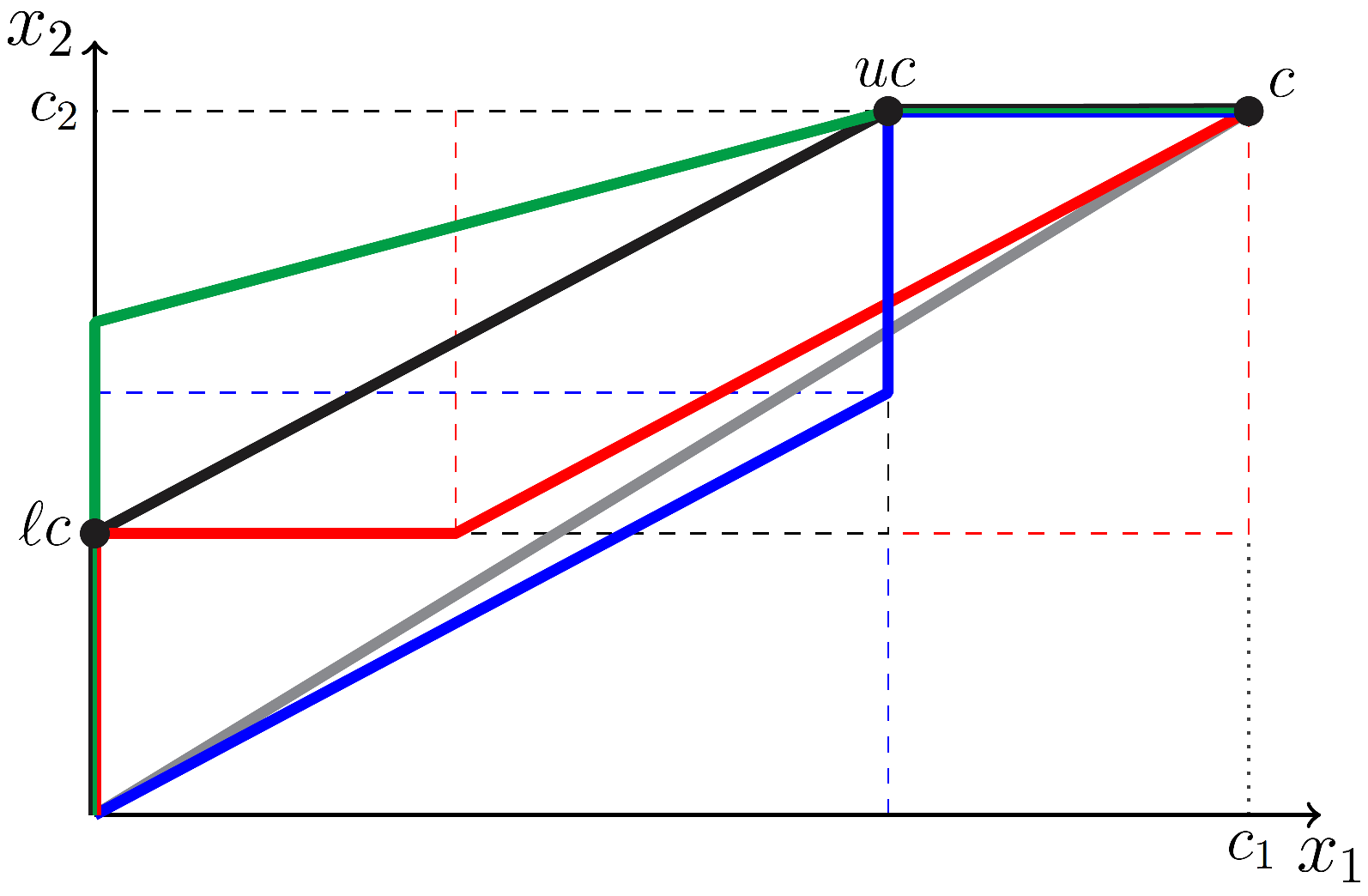}
  \caption{Independence of the axioms}
  \label{fig_th_indep}
  \end{figure}

We conclude by showing that the three axioms are independent. The black path in Figure \ref{fig_th_indep} shows the implementation of the exclusion dilation operator to the proportional rule (the benchmark rule is depicted in gray).

  \begin{enumerate}[1)]
    \item Drop \textit{full exclusion}. Take a rule defined by the following steps. First, it applies proportionality between the origin and the value of the exclusion space, that is $S(c,\ell,u)=s_{1}c_{1}+s_{2}c_{2}$. Second, the rule awards the agent with $\ell_{k}>0$ until the distribution reaches the upper exclusion threshold. Finally, the residual is allocated to the agent with $u_{k}>1$ (see the blue path in Figure \ref{fig_th_indep}). For all $(c,E,\ell,u)\in\mathcal{S}$ and $i=1,2$, \\
  \[
    \delta_{i}(c,E,\ell,u)=
    \left \{
      \begin{array}{cl}
        \frac{s_{i}c_{i}}{s_{1}c_{1}+s_{2}c_{2}}E  
        & \textnormal{if} \ E\leq S(c,\ell,u),
        \\[1.00ex]
        s_{i}c_{i}+\frac{\ell_{i}c_{i}}{L(c,\ell)}(E-S(c,\ell,u)) & \textnormal{if} \ E\in(S(c,\ell,u),U(c,u)),
        \\[1.00ex]
        u_{i}c_{i} & \textnormal{if} \ E=U(c,u),
        \\[1.00ex]
        u_{i}c_{i}+\frac{(1-u_{i})c_{i}}{C-U(c,u)}(E-U(c,u)) & \textnormal{if} \ E> U(c,u).
      \end{array}
    \right.
  \]
  Note that this function satisfies \textit{proportional exclusion invariance} with $m_{i}=0$, for each $i=1,2$.
    \item Drop \textit{null exclusion}. Take a rule defined by the following steps. First, it allocates the endowment according to the lower exclusion threshold. Second, the rule awards the agent with $u_{k}>1$ until the distribution reaches the amount $C-S(c,\ell,u)$. Finally, the residual is allocated in proportion to the exclusion thresholds (see the red path in Figure \ref{fig_th_indep}). For all $(c,E,\ell,u)\in\mathcal{S}$ and $i=1,2$, \\
  \[
  \small
    \delta_{i}(c,E,\ell,u)=
    \left \{
      \begin{array}{cl}
        \frac{\ell_{i}c_{i}}{L(c,\ell)}E  & \textnormal{if} \ E< L(c,\ell),
        \\[1.00ex]
        \ell_{i}c_{i} & \textnormal{if} \ E=L(c,\ell),
        \\[1.00ex]
         \begin{array}{c}
           \ell_{i}c_{i}
           \\
           +\frac{(1-u_{i})c_{i}}{C-U(c,u)}(E-L(c,\ell))
         \end{array}
         & \textnormal{if} \ E \in(L(c,\ell),C-S(c,\ell,u))),
        \\[3.00ex]
         \begin{array}{c}
           (1-s_{i})c_{i}
           \\
           +\frac{s_{i}c_{i}}{S(c,\ell,u)}(E-C+S(c,\ell,u))
         \end{array}
         & \textnormal{if} \ E\geq C-S(c,\ell,u).
      \end{array}
    \right.
  \]
  Note that this function satisfies \textit{proportional exclusion invariance} with $m_{i}=1-s_{i}$, for each $i=1,2$.
    \item Drop \textit{proportional exclusion invariance}. Take a rule defined by the following steps. First, it allocates the endowment according to the lower exclusion threshold. Second, it awards the agent with $\ell_{k}>0$ half of their claim in the exclusion space, that is, $s_{k}c_{k}/2$. Third, the rule applies proportionality between this point and the upper exclusion threshold. Finally, the residual is allocated to the agent with $u_{k}<1$ (see the green path in Figure \ref{fig_th_indep}). For all $(c,E,\ell,u)\in\mathcal{S}$ and $i=1,2$, \\
  \[
  \small
    \delta_{i}(c,E,\ell,u)=
    \left \{
      \begin{array}{cl}
        \frac{\ell_{i}c_{i}}{L(c,\ell)}E  & \textnormal{if} \ E< L(c,\ell),
        \\[1.50ex]
        \ell_{i}c_{i} & \textnormal{if} \ E=L(c,\ell),
        \\[1.50ex]
         \begin{array}{c}
           \ell_{i}c_{i}
           \\
           +\frac{s_{i}c_{i}}{S_{a}(c,\ell,u)}\frac{\ell_{i}c_{i}}{2L(c,\ell)} (E-L(c,\ell))
         \end{array}
         & \textnormal{if} \ E\in(L(c,\ell),L(c,\ell)+S_{a}(c,\ell,u)),
        \\[4.00ex]
        \ell_{i}c_{i}\left(1+\frac{s_{i}c_{i}}{2L(c,\ell)}\right)  & \textnormal{if} \ E=L(c,\ell)+S_{a}(c,\ell,u),
        \\[3.00ex]
         \begin{array}{c}
           \ell_{i}c_{i}\left(1+\frac{s_{i}c_{i}}{2L(c,\ell)}\right) 
           \\
           + \frac{s_{i}c_{i}\left(1-\frac{\ell_{i}c_{i}}{2L(c,\ell)}\right)}{S(c,\ell,u)-S_{a}(c,\ell,u)}
           \\
           *(E-L(c,\ell)-S_{a}(c,\ell,u)) 
         \end{array}
         & \textnormal{if} \ E\in(L(c,\ell)+S_{a}(c,\ell,u),U(c,u)),
        \\[6.00ex]
        u_{i}c_{i} & \textnormal{if} \ E=U(c,u),
        \\[1.50ex]
        u_{i}c_{i}+\frac{(1-u_{i})c_{i}}{C-U(c,u)}(E-U(c,u)) & \textnormal{if} \ E> U(c,u),
      \end{array}
    \right.
  \]
  where $S_{a}(c,\ell,u)=\sum_{k=1,2}s_{k}c_{k}\frac{\ell_{k}c_{k}}{2L(c,\ell)}$.
  \end{enumerate}


\bibliographystyle{chicago}
\bibliography{references}


\end{document}